\documentclass[preprint]{aastex}
\usepackage[T1]{fontenc}
\usepackage[latin1]{inputenc}
\usepackage[]{graphicx}

\usepackage{graphicx, natbib, color, bm, amsmath, epsfig}

\newcommand{\enzo}{{\small Enzo}}
\newcommand{\menzo}{{EnzoMHD}}

\newcommand{\zeus}{{\small ZEUS}}

\newcommand{\grid}{{\tt grid}}
\newcommand{\iph}{i + \frac{1}{2}}
\newcommand{\imh}{i - \frac{1}{2}}
\newcommand{\jph}{j + \frac{1}{2}}
\newcommand{\jmh}{j - \frac{1}{2}}
\newcommand{\kph}{k + \frac{1}{2}}
\newcommand{\kmh}{k - \frac{1}{2}}
\newcommand{\half}{\frac{1}{2}}
\newcommand{\quart}{\frac{1}{4}}
\newcommand{\tquart}{\frac{3}{4}}

\newcommand{\dbd}[2]{\frac{ \partial #1 }{ \partial #2} }
\newcommand{\curl}{\nabla \times}
\newcommand{\divb}{$\nabla \cdot {\bf B}$}
\newcommand{\divbo}{$\nabla \cdot {\bf B} = 0$}
\newcommand{\intd}[1]{\int_#1^{#1 + \Delta #1} }

\newcommand{\bld}[1]{\mathbf{#1}}
\newcommand{\dc}[1]{}

\newcommand{\mhdlabel}{{_{MHD}}}
\newcommand{\gravlabel}{{_{Grav}}}
\newcommand{\fclabel}{{_{fc}}}
\newcommand{\explabel}{{_{exp}}}

\begin{document}
\title {Cosmological AMR MHD with Enzo}
\author{D. C. Collins\altaffilmark{1}, H. Xu \altaffilmark{1,2}, M.L. Norman\altaffilmark{1} , H. Li\altaffilmark{2}, S. Li\altaffilmark{2}}
\altaffiltext{1}{Center for Astrophysics and Space Sciences, University of California, San Diego,
9500 Gilman Drive, La Jolla, CA 92093 }
\altaffiltext{2}{Theoretical division, Los Alamos National Lab, Los Alamos, NM 87545}
\begin{abstract}
In this work, we present \menzo, the extension of the cosmological
code \enzo\ to include the effects magnetic fields through the ideal MHD approximation.  We use a higher order Godunov Riemann solver
for the computation of interface fluxes.  We use two constrained transport methods
to compute the electric field from those interface fluxes, which simultaneously advances
the induction equation and maintains the divergence of the magnetic field.
A third order divergence free reconstruction technique is used to interpolate the magnetic
fields in the block structured AMR framework already extant in \enzo.
This reconstruction also preserves the divergence of the magnetic
field to machine precision.  We use
operator splitting to include gravity and cosmological expansion.  We
then present a series of cosmological and non cosmological tests
problems to demonstrate the quality of solution resulting from this
combination of solvers.
\end{abstract}



\maketitle
\section{Introduction}

Enzo is an adaptive mesh refinement (AMR), grid-based hybrid code
(hydro + N-Body) which is designed to do simulations of cosmological
structure formation. It uses the block-structured AMR algorithm of
\citet{Berger89} to improve spatial resolution where required, such as
in gravitationally collapsing objects. The method 
is attractive for cosmological applications because it: 1) is
spatially- and time-adaptive, 2) uses accurate and well-tested
grid-based methods for solving the hydrodynamics equations and 3) can
be well optimized and parallelized. The central idea behind AMR is to
solve the evolution equations on a fixed resolution grid, adding finer
grids in regions that require enhanced resolution. Mesh refinement can be
continued to an arbitrary level, based on criteria involving any
combination of (dark-matter and/or baryon) over density, Jeans length,
cooling time, etc, enabling users to tailor the adaptivity to the
problem of interest. Enzo solves the following physics models:
collisionless dark-matter and star particles, using the particle-mesh
N-body technique \citep{Hockney85}; gravity, using FFTs on the root
grid and multigrid relaxation on the subgrids; cosmic expansion; gas
dynamics, using the piecewise parabolic method (PPM) \citep{Colella84}
as extended to cosmology by \citet{Bryan95};
multi-species non-equilibrium ionization and $H_{2}$ chemistry, using
backward Euler time differencing \citep{Anninos97}; radiative heating
and cooling, using subcycled forward Euler time differencing
\citep{Anninos94}; and a parameterized star formation/feedback recipe
\citep{Cen93}. Enzo has been successfully used in many cosmological
applications, including star formation
\citep{Abel00,Abel02,O'Shea05,O'Shea07}, Lyman-alpha
forest \citep{Bryan99,Jena05}, interstellar
medium \citep{Kritsuk02, Kritsuk04} and galaxy clusters \citep{Bryan98, Loken02, Motl04,
  Hallman06}. More informations about Enzo are available at
http://lca.ucsd.edu/projects/enzo

One important piece of physics that is missing from this list is a
proper treatment of magnetic fields.  Magnetic fields have a broad
range of impacts in a broad range of physical situations, from galaxy
clusters to protostellar core formation.  Magnetic forces can shape
morphology of objects by forcing flow along the field lines.  They can
alter the energy balance by providing sources of pressure and energy.  They can
alter cooling rates by trapping electrons.  Alfven waves can
redistribute angular momentum throughout an object.  They create X-ray cavities seen in some
galaxy clusters.  They accelerate cosmic rays,
which play a crucial role in the energy balance of the galaxy and
galaxy clusters.  They also play a role in galactic star formation,
potentially removing angular momentum from collapsing objects and
launching protostellar winds.  Creating a functional cosmological MHD
code takes more than a single algorithm.  The purpose of this paper is to 
document the construction and performance of the algorithms that will be used in MHD simulations
with Enzo in the future, as well as simulations that have already been
done \citep{Xu08b,Xu08}

\menzo\ is also a purpose code.  In this paper,
we will discuss it as a cosmological code, but all the same
machinery applies in non-cosmological mode.  All algorithms used here
reduce to the non-cosmological limit by setting $a\rightarrow
1, \dot{a} \rightarrow 0,$ and $\ddot{a} \rightarrow 0$.  This
removes any frame dependent terms in the equations.  

We will describe the numerical procedures in
section \ref{sec.Numerics}, present test problems in
section \ref{sec.Tests}, and present conclusions and future plans in
section \ref{sec.Conclusion}. In appendix \ref{sec.Schematic} we
present a simplified schematic to unify the pieces of the solver, and
in appendix \ref{sec.recon} and \ref{sec.fc} we expand on some of
the more complex numerical procedures.

\section{Numerics}\label{sec.Numerics}
\subsection{ Cosmological MHD Equations } \label{sec.Equations}
	
\menzo\ solves the MHD equations in a comoving coordinate frame.
\begin{align}
\frac{\partial \rho}{\partial t} + \frac{1}{a} \nabla \cdot ( \rho {\bf v}) & =  0 \label{eq.Rho}\\
\frac{\partial \rho {\bf v}}{\partial t} + 
  \frac{1}{a} \nabla \cdot (\rho {\bf v v} + \bar{p} - {\bf B B}) 
  & =  - \frac{\dot{a}}{a} \rho {\bf v} - \frac{1}{a}\rho \nabla  \Phi  \label{eq.Momentum}   \\
\frac{\partial E}{\partial t} + 
 \frac{1}{a}\nabla \cdot [{\bf v}(\bar{p}+E)-{\bf B}({\bf B}\cdot {\bf v})]  
 & =  - \frac{\dot{a}}{a}(\rho v^{2}+\frac{2}{\gamma - 1}p+\frac{B^{2}}{2}) 
 - \frac{\rho}{a} {\bf v}\cdot \nabla \Phi \label{eq.Energy}\\
\frac{\partial {\bf B}}{\partial t} - 
 \frac{1}{a} \nabla \times ({\bf v} \times {\bf B}) 
 & =   -\frac{\dot{a}}{2a}{\bf B} \label{eq.Induction}
\end{align}
with the equation of state 
\begin{align}
E & = \frac{1}{2} \rho v^2 + \frac{p}{\gamma-1} + \frac{1}{2}B^2 \label{eq.EOS}\\
\bar{p} & = p + \frac{1}{2} B^2 \label{eq.TotalPressure}
\end{align}

Here, $\rho$ is the comoving density, p is the comoving gas pressure,
v is the proper peculiar velocity, ${\bf B}$ is the comoving magnetic
field, E is the total peculiar energy per unit comoving volume,
$\bar{p}$ is the total comoving pressure, $\gamma$ is the ratio of the
specific heats, $\Phi$ is the proper peculiar gravitational potential
from both dark-matter and baryons, $a=(1+z_i)/(1+z)$ is the expansion
factor and t is time. 

In this formulation, the comoving quantities that are evolved by the solver are
related to the proper observable quantities by the following
equations:

\begin{align}
\rho_{proper} & = \rho * a(t)^3 \label{eq.ProperRho}\\
p_{proper} & = p_{comoving}*a^3 \label{eq.ProperEnergy}\\
{\bf v}_{proper} & = {\bf v}_{comoving} - \dot{a}{\bf x} \label{eq.ProperVelocity}\\
\Phi_{proper} & = \Phi - \frac{1}{2} a \ddot{a} \vec{x}^2 \label{eq.ProperPhi}\\
{\bf B}_{proper} & = {\bf B}_{comoving} a^{\frac{-3}{2}} \label{eq.ProperBfield}
\end{align}

It should be noted that the relationship between ${\bf B}_{proper}$
and ${\bf B}_{comoving}$ in equation \ref{eq.ProperBfield} is
different than that stated in other cosmological MHD codes like
\citet{Li06}.  This is due to the additional expansion factor that we
use in equation \ref{eq.Induction}.  The proper magnetic field
decreases proportional to $a^{-2}$ in all formulations of the
cosmological MHD equations, but in the formulation we use one half
power of $a$ is included as a comoving source term and is due to the
redshifting of the photons that carry the magnetic field.  

For non-cosmological simulations, the same equations hold, but with with $a=1, \dot{a}=0$
and $\ddot{a} = 0$.  This effectively removed each appearance of $a$
from the left hand side, and eliminates the terms involving $\dot{a}$
from the right.  For ease of reference, these are:

\begin{align}
\frac{\partial \rho}{\partial t} + \nabla \cdot ( \rho {\bf v}) & =  0 \label{eq.Rho_nonCos}\\
\frac{\partial \rho {\bf v}}{\partial t} + 
  \nabla \cdot (\rho {\bf v v} + \bar{p} - {\bf B B}) 
  & =  - \rho \nabla  \Phi  \label{eq.Momentum_nonCos}   \\
\frac{\partial E}{\partial t} + 
 \nabla \cdot [{\bf v}(\bar{p}+E)-{\bf B}({\bf B}\cdot {\bf v})]  
 & =  - {\bf v}\cdot \nabla \Phi \label{eq.Energy_nonCos}\\
\frac{\partial {\bf B}}{\partial t} - 
 \nabla \times ({\bf v} \times {\bf B}) 
 & =   0
\end{align}

with the same equation of state, equations \ref{eq.EOS} and
\ref{eq.TotalPressure}.  Here, $\rho$ is the density, p is the gas pressure,
v is the velocity, ${\bf B}$ is the magnetic field, E is the total energy per unit volume,
$\bar{p}$ is the total gas pressure, $\gamma$ is the ratio of the
specific heats, $\Phi$ is the gravitational potential.  The mechanism
to switch between the two systems of equations will be described in
section \ref{sec.HyperbolicTerms}.

To solve these equations, we operator split eqns (\ref{eq.Rho})-(\ref{eq.Induction}) into four parts: the
left hand side of equations (\ref{eq.Rho})-(\ref{eq.Energy}), the left hand side of equation (\ref{eq.Induction}), the gravitational acceleration (the two terms
involving $\nabla \Phi$), and the expansion terms (the two terms
involving $\dot{a}$.)  These will be discussed in sections
\ref{sec.HyperbolicTerms} - \ref{sec.CT}.  In section
\ref{sec.DualEnergy}, we will discuss the dual energy formulation in
Enzo for hypersonic flows, and in section \ref{sec.AMR} we will
discuss the Adaptive Mesh Refinement algorithm.  We first discuss the
data structures used to carry all this data in section \ref{sec.data}

In the following, we will often have cause to separate the purely fluid
dynamical quantities $\rho, \vec{v}, E$ from the magnetic field
$\vec{B}$.  Unless otherwise noted, 'fluid quantities' will refer to
the former only.

For ease of reference, we have supplied a schematic summary of the steps
involved in appendix \ref{sec.Schematic}.

\subsection{Data Structure}\label{sec.data}

In \enzo, both parallelism and AMR are done in block decomposed
manner.  Each patch of space, called a \grid,  is treated as a unique computational
problem with Dirichlet boundary conditions which are stored in a
number of Ghost Zones (see section \ref{sec.BoundaryConditions}.)  The number of ghost zones depends on the
method used.  The pure-hydro methods in \enzo,  \zeus\ and PPM, use 3
ghost zones.  The method we describe here uses 5 ghost zones.  

\begin{figure}[p]
\includegraphics[width=1.0\textwidth]{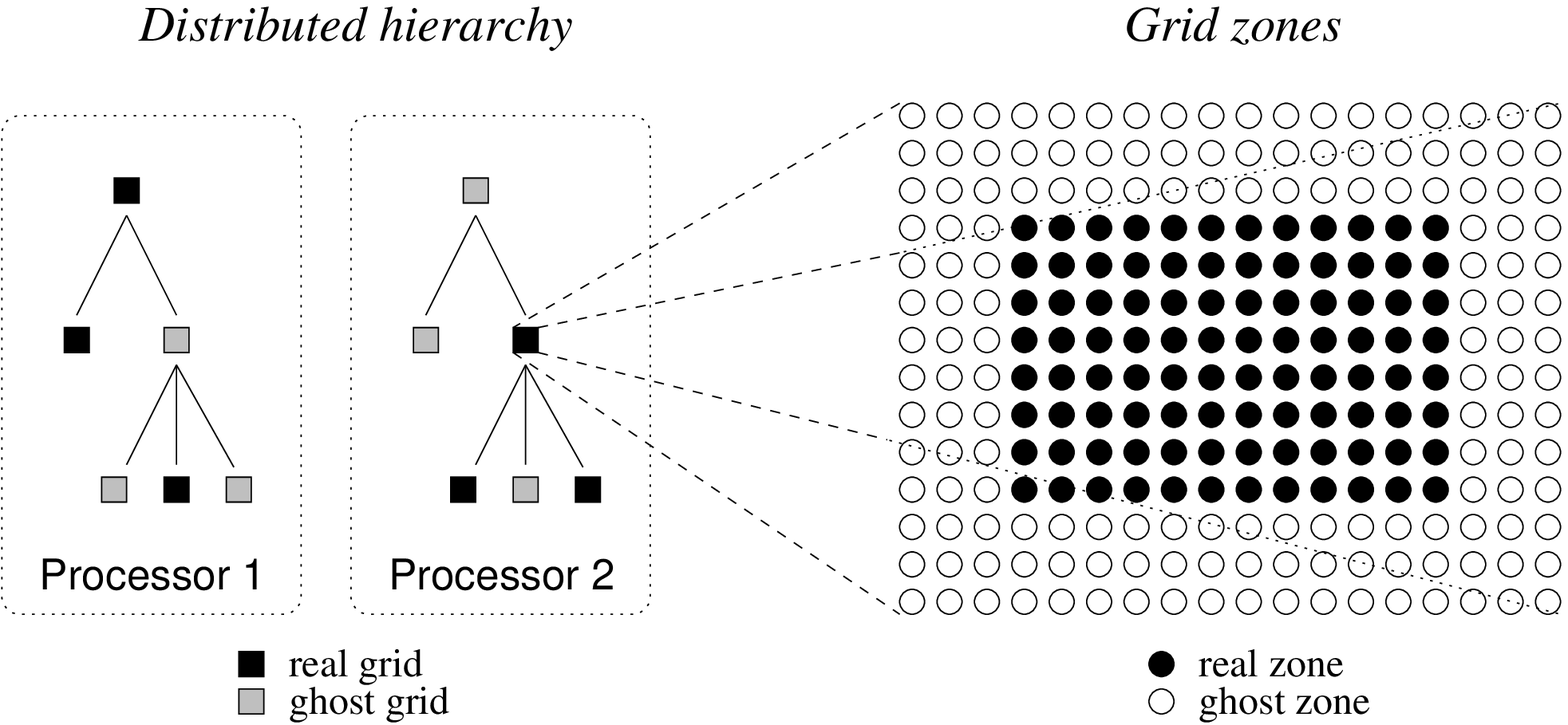}
\caption{A schematic of a parallel AMR hierarchy on two processors (left) and a grid patch with ghost zones (right).  Image courtesy James Bordner, initially appeard in \citep{Norman07}}
\label{fig.amr_hierarchy}
\end{figure}

Grids are arranged in a strictly nested hierarchy, with each grid
having a cell width half that of its parent (pure hydro \enzo\ can
take any integer refinement, but the interpolation for MHD is
restricted to factors of 2.)  See figure \ref{fig.amr_hierarchy}.  Each processor keeps a copy of the entire hierarchy, while only one of the processors actual stores the data.

For all physics modules described in this paper, an individual grid
cares not for where it sits in space or the hierarchy, and
communicates with other grids only through boundary condition fills
(section \ref{sec.BoundaryConditions}) and the AMR cycle (section \ref{sec.AMR}).

\menzo\ in its default mode tracks 14 fields, stored at 3 different
points of the cell.  The 5 hydrodynamic quantities, $\rho, {\bf v}, E_{total}$
are stored at the center of the cell, denoted $(i,j,k)$, and represent the volume average
of the respective quantities.  These are the same quantities stored in
non-MHD \enzo.  

\menzo\ tracks 2 copies of the magnetic field and the electric field.
One copy of the magnetic field is stored in the face of
the cell perpendicular to that field component, and represents the 
area average of that field component over that face.  This is the
primary representation of the magnetic field.  So $B_{f,x}$ is
stored in the center of the $x$ face, denoted $(\imh, j,k)$,
$B_{f,y}$ in the $y$ face at $(i,\jmh, k)$, and $B_{f,z}$ in the
$z$ face at $(i,j,\kmh)$.  It is this field that remains divergence
free under the cell centered divergence operator:
\begin{align}
\nabla \cdot {\bf B_f} = & \frac{1}{\Delta x} ( B_{f,x,\iph,j,k} -
B_{f,x,\imh,j,k}) + \nonumber \\
& \frac{1}{\Delta y} ( B_{f,y,i,\jph,k} - B_{f,y,i,\jmh,k}) + \label{eqn.divb}\\
& \frac{1}{\Delta z} ( B_{f,z,i,j,\kph} - B_{f,z,i,j,\kmh}) \nonumber
\end{align}
The magnetic data structures are one element longer in each longitudinal
direction, so for an $nx\times ny \times nz$ grid patch, the $B_{f,x}$
structure is $(nx + 1)\times ny \times nz$.

The second representation of the magnetic field is centered with the
fluid quantities at the center of the cell.  This field is used
wherever a cell centered magnetic quantity is needed, most notably in
the hyperbolic solver in section \ref{sec.HyperbolicTerms}.  It's
equal to the first order average of the face centered magnetic field:
\begin{align}
  B_{c,x,i,j,k}^{n+1} = 0.5*( B_{f,x,i+\half,j,k} + B_{f,x,i-\half,j,k} ) \nonumber\\
  B_{c,y,i,j,k}^{n+1} = 0.5*( B_{f,y,i,j+\half,k} + B_{f,y,i,j-\half,k} ) \label{eqn.center}\\
  B_{c,z,i,j,k}^{n+1} = 0.5*( B_{f,z,i,j,k+\half} + B_{f,z,i,j,k-\half} ) \nonumber
\end{align}

The final data structure used in \menzo\ is the Electric Field, which
is stored along the edges of the computational cell.  This represents
a linear average of the electric field along that line element.
Each component is centered along the edge its parallel to, so $E_x$
lies along the $x$ edge of the cell at $(i,\jmh,\kmh)$, etc.   It is longer
than the fluid fields by one in each transverse direction, so $E_x$
would be $nx \times (ny + 1 ) \times (nz + 1)$.

Each grid also stores one copy of each of the above mentioned fields
for use in assigning ghost zones to subgrids.  This is described
further in \ref{sec.BoundaryConditions}.  A temporary field for fluxes
is also stored, which exists only while the hyperbolic terms are being
updated.  This data structure is also stored on the faces of the
zone.  There are three fluxes for all 7 MHD quantities. 

For other configurations of \menzo, more or fewer fields may be used.
In purely isothermal mode (which is at present an option only in
\menzo, not in \enzo) the total energy field is not tracked, and the
isothermal sound speed is taken as a global scalar quantity.  This
reduces the number of fields tracked everywhere the total energy shows
up.  With dual energy formalism on (see section \ref{sec.DualEnergy})
an additional field corresponding to either gas energy or entropy is
stored, giving an additional field where needed.  Future work will include 
multi-species chemistry and more complex cooling, which will include additional 
fields for each species.

\subsection{Consistency} \label{sec.consistency}
In several places throughout the flow of \enzo, there may be more than
one data structure using and writing to a given variable at a given
point in space.  Ghost zones and face centered fields (fluxes and
magnetic fields) are examples of this.  In \menzo, it is imperative that
all data at a given point is identical, regardless of the data structure
describing it.  This may seem like an unnecessary comment, but it
isn't;  in pure hydro simulations, numerical viscosity will damp out
small perturbations caused by slight inconsistencies in data
description. Thus in practice, especially in large, stochastic simulations,
errors can  go unnoticed.  Often these discrepancies are negligible,
other times not, especially when one is concerned with the
conservation of a particular variable, like \divb.  By construction \menzo\ preserves \divb\ to
machine precision, but it never \emph{forces} \divbo; so if it's
not zero at the beginning of a time step, it's not going to be at the
end, either.   It is also worth mentioning that inconsistencies in any
quantity will cause inconsistencies in the flow, which will in turn
cause \divb\ issues.  Thus any improper handling of \emph{any} fluid
quantity will cause errors in \divb\ that will persist and usually
grow to catastrophic proportions in a relatively short period of time. 

\begin{figure}[ht]
\centering
\epsfig{file = 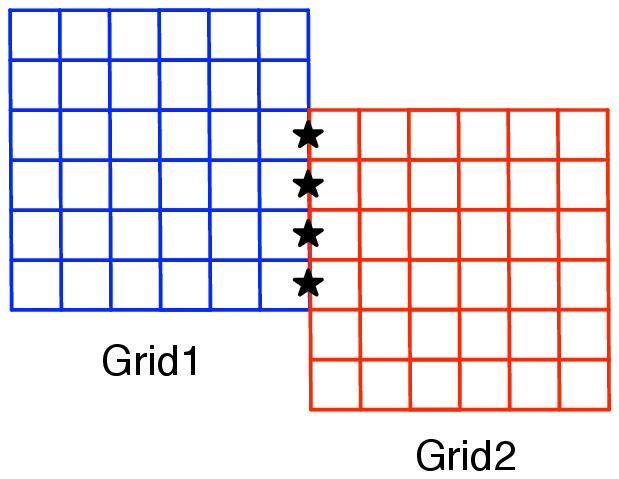, width=2in}
\caption{ Data redundancy of the face centered magnetic fields: the
face centered field denoted by the stars are updated by both grid 1
and grid 2.  Enough ghost zones are exchanged to ensure that the
entire stencil for the update of these fields is the same in both data
structures.  
\label{fig.overlap}}

\end{figure}

There is a prominent redundancy in the magnetic field, namely the
field on the surface of the active zones of grids.  See figure
\ref{fig.overlap}.  Care is taken to
include enough ghost zones, and frequent enough ghost zone exchange
between grids, that after a time step, two neighboring grids have
reached exactly the same answer on the surface between the two grids
completely independently.

\subsection{Time Stepping} \label{sec.dt}
\begin{figure}[p]
\includegraphics[width=1.0\textwidth]{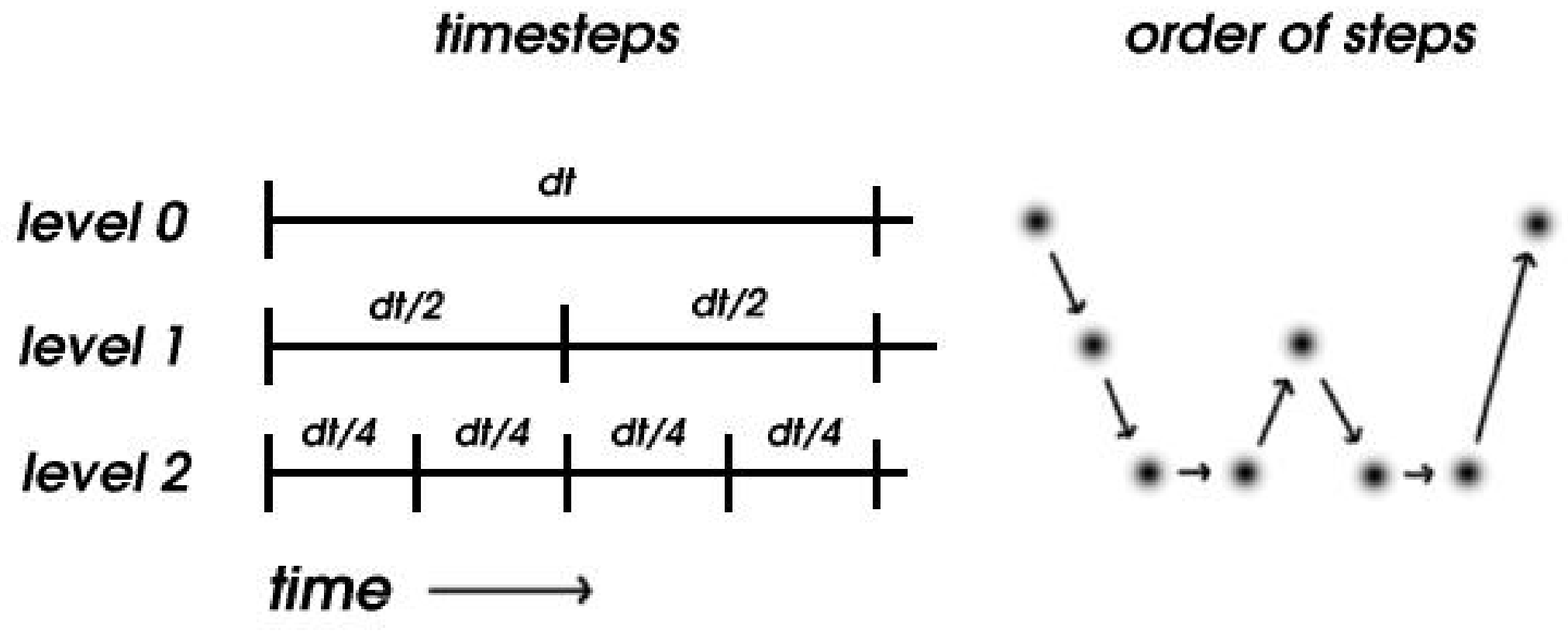}
\caption{ A depiction of the timestep strategy in \enzo}
\label{fig.timesteps}
\end{figure}

Enzo uses hierarchical time stepping to determine it's time step.  The
minimum of 4 different criteria is taken for each level, which will be
described in sections \ref{sec.dt_cfl} - \ref{sec.dt_particles}.
Timesteps are taken in order of coarsest to finest, in a 'W' cycle.
See figure \ref{fig.timesteps}. 
Given 3 levels, level 0 takes the first step of $\Delta t$.  Then
level 1 takes a single step of $\Delta t/2$.  Then level 2 takes one
step of $\Delta t/4$.  Then, given that there are only three levels, it
takes another timestep so it is temporally in line with the level above.  
The last three steps repeat: level 1 then takes its
second and final step of $\Delta t/2$ so it is now at the same time as
level 0, followed by two steps on level 2.  

In principle, if a given level has a cell size $\Delta x$ and the
next level of refinement has cell size $\frac{\Delta x}{r}$, where $r$ is the refinement factor, the more
refined grid will have, in principle, time step size $\frac{\Delta t}{r}$.  In \enzo, the
step size is chosen for each level and each subgrid time step.  In
practice, owing to more finely resolved structures having slightly
higher fast shock speeds, fine grids may in fact take more than $r$ time steps for
each parent grid step.  In some rare cases, such as cosmological expansion limiting, a finer grid 
may take less than $r$ steps.

\subsubsection{Time Stepping: Hydro} \label{sec.dt_cfl}
For the hydrodynamics, the harmonic mean of the 3 Courant conditions is used.
This was demonstrated to be the most robust time stepping criterion
possible for multi dimensional flows by \citet{Godunov61}.
\begin{align}
\Delta t_{hydro} = & \frac{1}{1/t_x + 1/t_y + 1/t_z} \nonumber \\
t_x = & min(\frac{\Delta x}{c_{f,x}}) \\
t_y = & min(\frac{\Delta y}{c_{f,y}}) \nonumber\\
t_z = & min(\frac{\Delta z}{c_{f,z}}) \nonumber
\end{align}
where the $min$ is taken over the zones on a level, and $c_{f,x},c_{f,y}$ and $c_{f,z}$ are the fast MHD shock speeds along each axis:
\begin{equation}
c_{f,x}^2 = \half \left(a^2 + \frac{\bold{B}\cdot\bold{B}}{\rho} + 
\sqrt{(a^2 + \frac{\bold{B}\cdot\bold{B}}{\rho})^2 - 4 a^2 B_x^2/\rho}\right)
\end{equation}
and similar definition for the other two.

\subsubsection{Time Stepping: Gravitational
  Acceleration}\label{sec.dt_gravity}
The time step is also restricted to be less than the time it takes for
the gravitational acceleration alone to move a parcel of fluid half of
one zone.
\begin{equation}
\Delta t_{accel} = min( \half sqrt{\frac{\Delta x}{a_i}} )
\end{equation}
where $i=x,y,z$ and the $min$ is taken of the zones on a level.

\subsubsection{Time Stepping: Cosmological
  Expansion} \label{sec.dt_expansion}
An additional restriction comes from the cosmological expansion,
requiring the timestep to be less than the cosmological expansion
timescale,
\begin{equation}
\Delta t_{expansion} = \eta \frac{a}{\dot{a}}
\end{equation}
where $\eta$ is typically 0.01.
\subsubsection{Time Stepping: Particle Motion}\label{sec.dt_particles}
The fourth timestep criterion is based on restricting particle
displacement in a single timestep to be smaller than a single zone:
\begin{equation}
\Delta t_{particles} = min(\frac{a \Delta x}{v_{i,p}})
\end{equation}
where $min$ is over velocity component $i$ and particle $p$.

\subsection{Boundary Conditions and Ghost Zones}\label{sec.BoundaryConditions}

Ghost Zones are filled in one of three means.  
\begin{enumerate}
  \item {\bf Copying.}  The dominant mechanism for filling ghost zones
copying from active zones that occupy the same physical space.  This also
takes into account periodic boundary conditions.  For
\menzo, face centered fields are copied from the faces of all cells,
including those that border on active cells.  This is somewhat
redundant for reasons described in \ref{sec.consistency}.
  \item {\bf External} Root grids that lie along the domain wall 
filled with the external boundary routine.    If the
external boundary condition is not periodic, the grids zones are
filled by a predetermined algorithm;  for instance, outflow boundary
conditions set ghost zones to be equal to the outermost active zone,
akin to a  Neumann condition of zero slope.   These involve outflow,
reflecting, and a completely general 'inflow'.  Note that this is
called only on the root grid, and not on subgrids that happen to lie
on the edge.   This can cause spurious waves at reflecting or outflow
boundaries with AMR.  Also note for \menzo, the only external boundary
conditions that have been tested are periodic and outflow.
 \item {\bf Interpolation} The third mechanism is
used on refined grids whose ghost zones do not occupy the active space
of another grid;  these grids have their ghost zones filled by
interpolation from the parent grid.  Since \enzo\ uses hierarchical time stepping,
subgrid steps that begin in the middle of a parent grid step fill
their ghost zones from a linear interpolation of the parent grid
time steps at $t^n$ and $t^{n+1}$.
\end{enumerate}

\subsection{Left Hand Side: Hyperbolic terms}\label{sec.HyperbolicTerms}

With the exception of the $1/a$ term that appears in front of each
$\nabla \cdot$ operator, the left hand side of equations
\ref{eq.Rho}-\ref{eq.Induction} are the familiar 
Ideal MHD equations.  A form of equations (\ref{eq.Rho}) -
(\ref{eq.Induction}) more relevant for this treatment is the following:
\begin{equation}
 \label{eq.CL} 
\frac{\partial {\bm V}}{\partial t} + \frac{\partial{\bm F}}{\partial x} =
 0
\end{equation}
where
\begin{equation}
\label{eqn.q}
{\bf V} = \begin{pmatrix}
\rho \\
\rho v_x \\
\rho v_y\\
\rho v_z\\
B_y\\
B_z\\
E\end{pmatrix} \\
\end{equation}
\begin{equation}
\label{eqn.F}
{\bf F} = \begin{pmatrix}
\rho v_x \\
\rho v_x^2 + p + B^2/2 - B_x^2\\
\rho v_x v_y - B_x B_y \\
\rho v_x v_z - B_x B_z \\
B_y v_x - B_x v_y = -E_z\\
B_z v_x - B_x v_z = E_y\\
(E + p + B^2/2) v_x - B_x( {\bf B}\cdot{\bf v} )
\end{pmatrix} \\
\end{equation}
\begin{equation}
p = (E-\half \rho {\bf v}^2 + \half {\bf B}^2 (\gamma - 1) ) \\
\end{equation}
These form a hyperbolic system of equations, which
have been studied extensively in the literature.  To take advantage of
the work already done on this type of system of equations for our 
cosmological algorithm, we first
multiply the cell width $dx$ by the expansion factor $a$.   
This allows us to use any non-cosmological solver for cosmological applications.
Upon completion of the solver, $dx$ is divided by $a$ to restore $dx$ to the original comoving
value.  

Equation \ref{eq.CL} is solved by first re-writing it in conservation
form, that is taking suitable integrals in time and space.  The
resulting update is, in one dimension,
\begin{align}
  \hat{V}^{n+1}_{i,j,k} = \hat{V}^n_{i,j,k} - 
  \frac{\Delta t}{\Delta x}( \hat{F}^{n+\half}_{x,\iph,j,k} - \hat{F}^{n+\half}_{x,\imh,j,k} ) 
\end{align}
where $\hat{V}$ represents the spatial average of the
conserved quantities, and $\hat{F}$ represents an space and time average
of the flux, centered in time at $t = t+\Delta t/2$.  $\hat{V}$ is the
quantity we store in the cells, and $\hat{F}$ comes from the hyperbolic solver.

The solver we use to solve the hyperbolic equations is 
that of \citet{Li06}, which is comes in three parts: spatial
reconstruction, time centering, and the solution of the Riemann
problem.  Spatial reconstruction is done using piecewise linear
monotonized slopes on the primitive variables $(\rho, {\bf{v}}, p,
{\bf B})$.  Time centering of the interface states by $\Delta t/2$ 
is performed using either the MUSCL-Hancock \citep{Li06} or Piecewise Linear Method \citep{Colella86}
integration.  The Riemann problem is then solved using either the HLLC Riemann solver of
\citet{Li05},  HLLD solver of \citet{Miyoshi05}, or the isothermal
HLLD solver of \citet{Mignone07}.  These fluxes are computed for the
conserved, cell centered variables $(\rho, \rho {\bf v}, E, {\bf B_c})$. These fluxes
are then differenced to obtain the update
values of the fluid quantities only.  The fluxes for the magnetic
field are stored for use in the Constrained Transport algorithm, discussed in
section \ref{sec.CT}.  This is done in one dimension on
successive sweeps along the $x, y,$ and $z$ directions.  To reduce
operator splitting error, the order of the sweeps is permuted.  For
more details, see \citet{Li06}.    

In isothermal mode, the same method is used, but the energy terms in
$V$ and $F$ are removed, and only the isothermal HLLD can be used.

\subsection {Constrained Transport and the Divergence of {\bf B}} \label{sec.CT}	
One of the biggest challenges for an MHD code is to maintain the
divergence free constraint on the magnetic field ($\nabla \cdot \bf{B} =
0$). \citet{Brackbill80} found that non-zero divergence can grow
exponentially during the computation and cause the Lorentz force to be
non-orthogonal to the magnetic field.  There are three major ways to assure the
divergence remains zero. The first is a divergence-cleaning (or Hodge Projection)
approach by \citet{Brackbill80}, which solves an extra Poisson's
equation to recover $\nabla \cdot \bf{B}=0$ at each time step. But
\citet{Balsara04} found that non-locality of the Poisson solver
introduces substantial spurious small scale structures in the
solution.  Additionally, solving Poisson's equation on an AMR mesh is
computationally expensive.  The second method involves extending the MHD
equations to include a divergence wave, as done by  \citet{Powell99}, \citet{Dedner02}, which then advects the
divergence out of the domain.  As most of our solutions are done on
periodic domains, this is also an undesirable solution.  The third
method, and the one we have employed in \enzo, is 
the constrained transport (CT) method of \citet{Evans88}.  This method centers
the magnetic field on the faces of the computational cells and the
electric field on the edges.  Once the electric field is computed
(more on this later) it's curl is taken to update the magnetic field.
This ensures \divbo\ for all time, provided it's true initially.
\begin{align}
 \hat{B}^{n+1}_{f,x,\imh,j,k} = \hat{B}^n_{x,\imh,j,k} - \Delta t
 (& \frac{1}{\Delta y} (\hat{E}_{z,\imh,\jph,k} -
 \hat{E}_{z,\imh,\jmh,k} ) + \\
 & \frac{1}{\Delta z} (\hat{E}_{y,\imh,j,\kph} -
 \hat{E}_{y,\imh,j,\kmh} ) )  \nonumber
\end{align}

Plugging equation \ref{eqn.inductiond} into the divergence operator
\ref{eqn.divb} to find $\nabla \cdot {\bf B}_f^{n+1}$, one finds all
terms are eliminated except the initial divergence  $\nabla \cdot {\bf B}_f^{n}$.  

The CT algorithm of \citet{Evans88} was extended to work with finite
volume methods by \citet{Balsara99}.  This method uses the fact that the MHD Flux
has the electromotive force as two of
its components (see the $5^{th}$ and $6^{th}$ components of
eqn. \ref{eqn.F}), so using these components then incorporates all the
higher order and shock capturing properties of the Godunov solver into
the evolution of the electric field.  These components, which are centered at the face
the computational cell, are then averaged to obtain an electric field
at the edges of the cell.  This was the first CT method applied to
\enzo, so unless otherwise noted, the simulations presented here were
done with this method. The reader is encouraged to read \citet{Balsara99} for the
full details.  

\citet{Gardiner05} extended this idea to include higher
order spatial averaging, which eliminates a number of numerical
artifacts present in \citet{Balsara99} and increases the accuracy of
the method.  This method uses the fluxes from the Riemann solver, plus
additional information from the data in the cell to construct a linear
interpolation from the cell face to the cell edge.  The reader is
encouraged to see that paper for the details.

After the curl is taken and the face centered field ${\bf B_f}$ is
updated, it is then averaged to obtain ${\bf B_c}$, via equation \ref{eqn.center}.

\subsection{Right Hand Side: Gravitational Acceleration}\label{sec.GravityTerms}

In cosmological simulations, \enzo\ tracks the proper peculiar
gravitational potential.  
\begin{equation}
  \nabla^{2} \Phi = \frac{4\pi G}{a} (\rho_{b}+\rho_{d}-\rho_{0})
\end{equation}
where $\rho_{b}$ and $\rho_{d}$ are baryonic and dark matter comoving
density respectively, and $\rho_{0}$ is the comoving background
density. For non-cosmological simulations, the dark matter and
background density are ignored. 

The gravitational potential $\Phi$ is solved in Enzo using a
combination of methods.  First, the root grid potential (which
covers the entire computational domain) is solved for using  a fast
Fourier transform.  Then the subgrids (which hopefully do not cover the
computational domain) are solved using a multigrid relaxation
technique.  This resulting potential
$\Phi$ is then differenced to obtain the acceleration ${\bf g} = {\bf
  \nabla} \Phi$.  Specifically, 
\begin{equation}
{\bf g}_i = \half(\Phi_{i+1} - \Phi_{i-1})
\end{equation}

As mentioned before, the fluxes are computed at the half time point
$t+1/2 \Delta t$.  In order to keep the velocity and consistent
with this time centering, they are first advanced by a half
time step:
\begin{equation}
{\bf v} = {\bf v} + \frac{\Delta t}{2} {\bf g}
\end{equation}

After the fluxes are differenced to obtain the new state $v_x^{n+1}$,
these states are then updated with the accelerations.  For the
velocity update, a density field centered in time is used.  We follow
the same formulation used by \citet{Colella84}
\begin{align}
{v_x}^{n+1} & = v_x^{\prime n+1} + \Delta t \frac{\half (\rho^{n+1} + \rho^n) A_x}{\rho^{n+1}} \\
E^{n+1} & = E^{\prime n+1} - \half \rho^{n+1} ({v_x}^{\prime n+1})^2 + \half \rho^{n+1} ({v_x}^{n+1})^2 
\end{align}

\subsection{Right Hand Side: Expansion Source Terms}\label{sec.ExpansionTerms}

The cosmological expansion source terms are treated
in much the same way as the gravitational source terms. First,
a half time step is added to the values before the flux is computed.

\begin{align}
\bold{v}^{\prime n} = &\bold{v}^n - \half \Delta t \frac{\dot{a}}{a} \rho^n\\
p^{\prime n} = & p^n - \half \Delta t \frac{\dot{a}}{a} 3(\gamma - 1) p^n\\
\bold{B}_c^{\prime n} = & \bold{B}_c^n - \half \Delta t \frac{\dot{a}}{2 a} \bold{B}_c^n
\end{align}

The quantities $\bold{v}^{\prime n}$,  $p^{\prime n}$ and
$\bold{B}^{\prime n}$ are then used
in the rest of the solver described in section
\ref{sec.HyperbolicTerms}. After the fluxes are differenced, the 
source terms are then added to the
fluid quantities in full. This is done in a semi-implicit manner, by
averaging the quantities to be updated in time.  For instance, the
expansion contribution to the magnetic field is
\begin{align}
\frac{\partial {\bf B}}{\partial t} =  -\frac{\dot{a}}{2a}{\bf B}
\end{align}
which is discretized
%
\begin{align}
{\bf B}_\explabel^{n+1} - {\bf B}^{n+1} =  -\frac{\dot{a}}{2a}(\frac{{\bf
    B}_\explabel^{n+1} + {\bf B}^{n+1}) }{2})
\end{align}
and solving for ${\bf B}_\explabel^{n+1}$ we have
\begin{align}
x = & \frac{\dot{a}}{4a} \\
{\bf B}_\explabel^{n+1} = &\frac{(1-x)}{(1+x)} {\bf B}^{n+1}
\end{align}
Pressure and velocity are updated in a similar manner.  See appendix \ref{sec.Schematic} for the full update.

\subsection{Dual Energy Formalism} \label{sec.DualEnergy}

Hypersonic flows are quite common in cosmological simulations.  Due to
the extremely large gravitational forces, the ratio of kinetic energy
$E_{kinetic}$ to gas internal energy $E_{internal}$ can be as high as $10^8$.  This
leads to problems when computing the internal energy in this type of
flow, as the universe does math with infinite accuracy, but computers
do not.  Higher order Godunov code typically track only the total
energy (equation \ref{eq.EOS}). Thus finding the internal energy from the total energy 
tracked by the software,
\begin{equation*}
E_{internal} = E_{total} -E_{kinetic}-E_{magnetic}
\end{equation*}
involves the small difference of two (or three) large numbers, which
causes problems when the small number ($E_{internal}$) is near the roundoff noise of
the original numbers ($E_{total}$ and  $E_{kinetic}+E_{magnetic}$).

To overcome this, we have implemented two algorithms that solve an
additional equation to track the small numbers; the modified entropy equation
given in \citet{Ryu93} and the internal energy equation given in
\citet{Bryan95}. These two  equations are: 
\begin{eqnarray}
\frac{\partial{S}}{\partial{t}} + \frac{1}{a} \nabla \cdot (S\bf{v}) & = & - \frac{3(\gamma -1)\dot{a}}{a}S \label{eq.Entropy}\\
\frac{\partial{\rho e}}{\partial{t}} + \frac{1}{a} \nabla \cdot (\rho e \bf{v}) & = & - \frac{3(\gamma-1)\dot{a}}{a}\rho e + \frac{p}{a} \nabla \cdot \bf{v}  
\end{eqnarray}
where $S\equiv p/\rho^{\gamma-1}$ is the comoving modified entropy and
$e$ is the internal energy. The modified entropy equation is valid
only outside the shocks where the entropy is conserved.   Use of
either (not both) of these equations is at the discretion of the simulator.

Through the course of the simulation, the ratio of internal energy to
total energy is monitored.  When this ratio is less than some preset
value $\eta$, one of the modified equations is used.  As in \citet{Li06}, we
use $\eta = 0.008$.  They note that reducing this parameter will cause a decrease in the volume filled by low
temperature gas, as most of the gas affected by the switch is cold, high velocity gas.  The optimal choice for this parameter 
is still an open question for the general situation.  \citet{Li06} compared this two approaches and found
  almost identical results.

\subsection{Adaptive Mesh Refinement} \label{sec.AMR}

Structured AMR, initially devised by \citet{Berger89}, is a technique
for increasing resolution of a simulation in parts of a simulation
that require higher resolution for increased accuracy or suppression
of numerical artifacts, while conserving memory and CPU cycles in
areas that don't.  Refinement criteria will not be described here, as
they vary from simulation to simulation.  AMR has four basic necessary
parts: 
\begin{enumerate}
 \item {\bf Patch Solver} This is the algorithm that actually solves
 the finite volume PDEs in question, as described by sections
 \ref{sec.HyperbolicTerms} - \ref{sec.DualEnergy}.  The approximations
 used for the patch solver are conservative in a finite
 volume sense, and the rest of the choices are made to preserve that conservation. 
 \item {\bf Refinement Operator} This is the routine that creates fine
 resolution elements from coarse ones.  In \enzo, we use conservative,
 volume weighted interpolation for the fluid quantities $\rho, E,
 \vec{v}$.  For the magnetic fields, we use the 
 method described by \citet{Balsara01}, with some slight
 modifications in implementation.  This method constructs a quadratic divergence free polynomial, and 
 area-weighted averages are used for the fine grid quantities. This is
 described in more detail in appendix  \ref{sec.recon}.
 \item {\bf Projection Operator} This is the routine that projects the
 fine grid data back to the parent coarse grid.  For \enzo, the parent
 grid is simply replaced by a volume-weighted average of the fine
 cells.  For the face centered magnetic field, this is an area
 weighted average, though in practice we don't explicitly average the
 magnetic field, as discussed in below and in appendix \ref{sec.mhdrecon}
 \item {\bf Correction Operator}  Once the projection operator
   replaces the solution on the coarse grids, the evolution on the
   coarse grids is no longer consistent with the underlying equations
   in the manner they were discretized.  That is to say, the total
   change of any conserved quantity inside the region is no longer
   equal to the flux across its surface.  For the \enzo\ hydro fields,
   this is corrected with the flux correction mechanism.  More details on this 
   and the modifications in \menzo\, see appendix \ref{sec.fc}
\end{enumerate}

\menzo\ does all of these steps for the fluid quantities, but for the
magnetic field it slightly alters this procedure.  In order to
overcome a shortcoming in the original data structures used in \enzo,
we combined the projection and correction operations for the magnetic
fields in one step.  The
net effect of the correction operator is to ensure that all zones are
updated by finest resolution fluxes available, even if they were updated
by coarse data initially.  For the magnetic field update, we don't
project the actual magnetic field that is of interest, but rather the
electric field (effectively the 'flux' for $B_f$), then take the curl
of the newly projected electric field.  Thus the coarse magnetic data
co-located with the fine grids get updated with the fine data, and the
bounding zones don't need correction at all.  

More detail on this process can be found in appendices \ref{sec.recon} and \ref{sec.fc}


\section {Numerical Experiments}\label{sec.Tests}

\menzo\ has many configurations available.  Here, we test some of the
possible configurations, to indicate the quality of solution possible
with \menzo.

\subsection {MHD Tests without AMR}
	
We first test our code in unigrid (fixed resolution) mode,
in order to ensure consistency of the patch solver with the algorithm
described in \citet{Li06}. We do two one dimensional cosmology tests
(Caustics and Zel'dovich Pancake), two one dimensional
non-cosmological tests (Brio and Wu and the Kim Isothermal), one 2d
non-cosmological test (Orszag Tang) and one 3d cosmological test.  

\subsubsection{Brio and Wu shock tube}\label{sec.BW}
The shock tube defined by \citet{Brio88} is a standard test of any
MHD solver, as it displays a
number of the important MHD waves, including a compound wave.
Compound waves are not a property of pure hydrodynamics, because the
system is convex. However, due do the more complex nature of the
MHD equations, certain initial conditions can cause flows in which at
one point the shock speed in a given family
is higher than the wave speed for that family, causing a
shock, but lower in the post shock region, causing a rarefaction
immediately following the shock.  

\begin{figure}[p]
\plotone{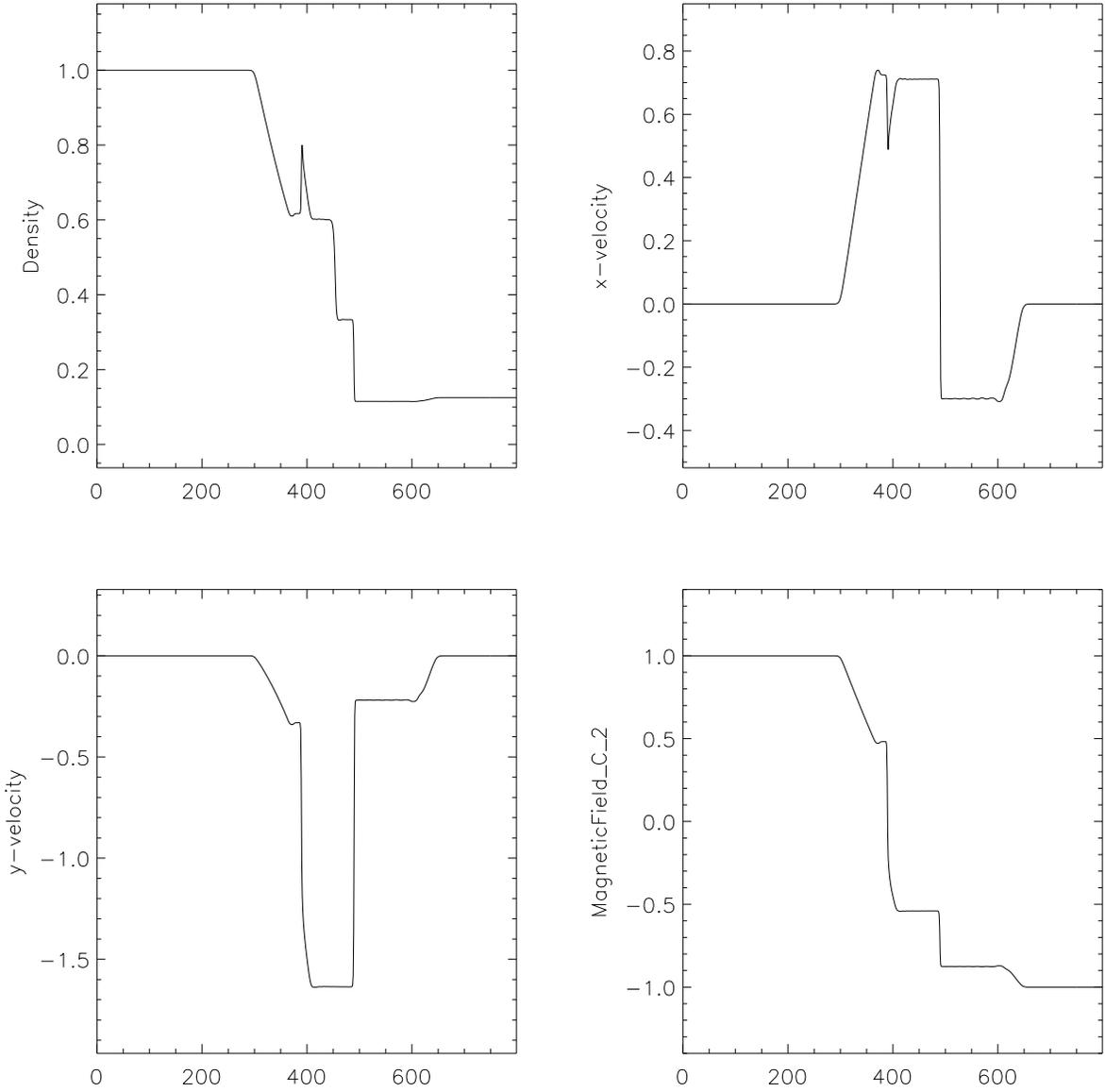}
\caption{The shock tube of \citet{Brio88}, showing from left to right
  a fast rarefaction, slow compound (shock+rarefaction), contact, slow
shock, and fast rarefaction. T=0.08, and 800 zones
were used.}
\label{fig.bw}
\end{figure}

This can be seen in figure \ref{fig.bw}.  The problem was run with
800 zones to a time $t=0.2$, using the HLLD solver in Enzo.   This shock
tube shows, from left to right, a fast rarefaction, slow compound
(shock+rarefaction), contact, slow shock, and fast rarefaction.  It
can be seen that this solver captures this shock tube problem quite
well.  

\subsubsection{Isothermal Tests}\label{sec.Kim}

One of the primary application areas of \menzo\ will be in
simulating turbulence and star formation in cold molecular clouds.
Due to the fast cooling time of these environments, an isothermal
equation of state is a good approximation a large portion of these
processes.  In simulations done by 
\citet{Kritsuk07} using \enzo\ and other works by the same authors an
isothermal equation of state is approximated by using an adiabatic
solver and setting $\gamma = 1.001$.  

\begin{figure}[p]
\plotone{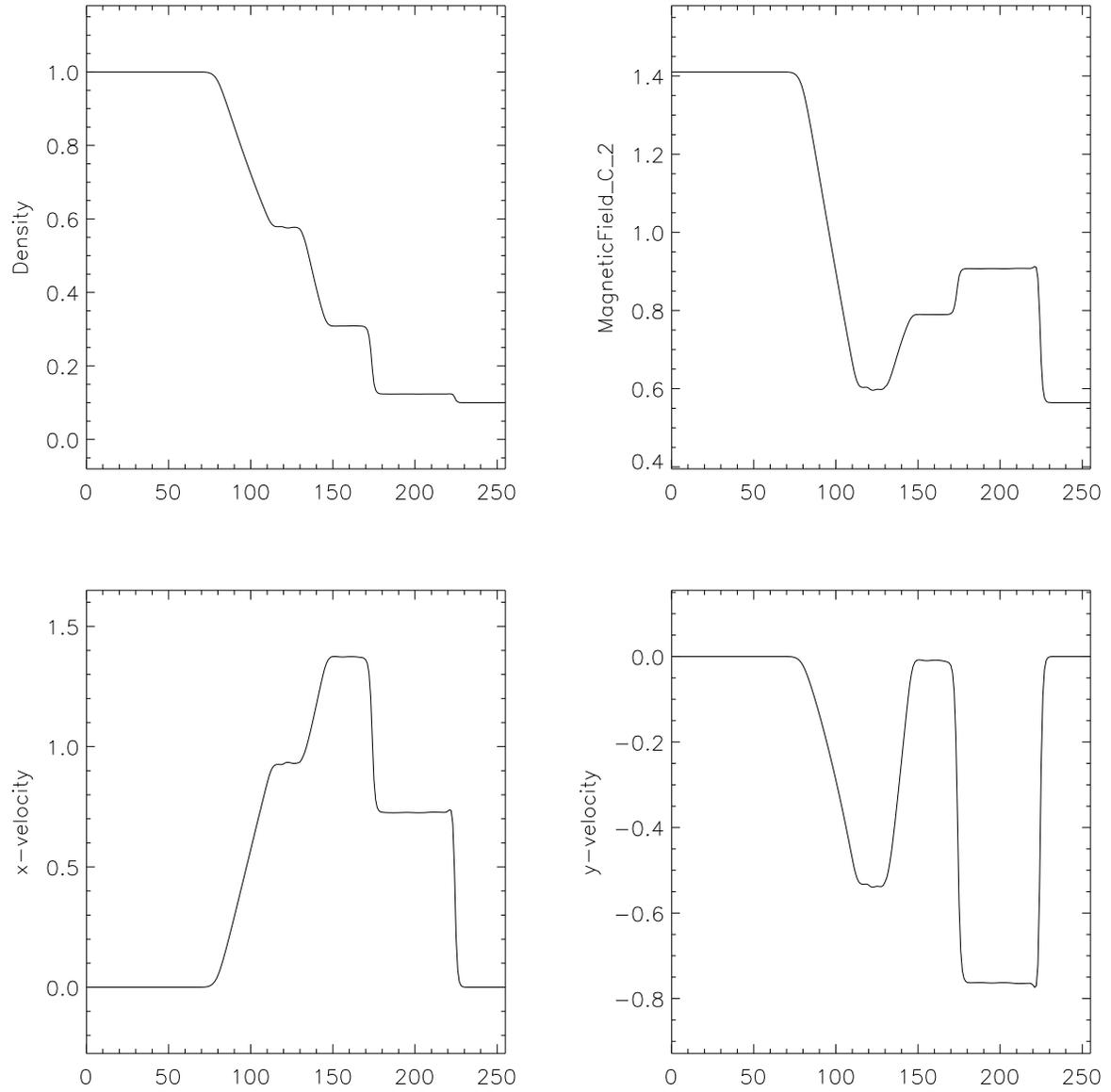}
\caption{The shock tube of \citet{Kim99}, run with 256 zones to t=0.1.}
\label{fig.kim_ad_hlld}
\end{figure}

To test if this approximation is appropriate for this code, we ran the
isothermal shock tube of \citet{Kim99}.  One can see from figure
\ref{fig.kim_ad_hlld} that this approach works well, as shock jumps
and positions are all correct, and features are reasonably sharp. This
test was run with 256 zones to a time of 0.1.

However, in turbulent simulations with gravitational collapse, the
measured value of the sound speed, $\sqrt{p/\rho}$, is initially uniform, but after a few hundred
timesteps can vary by as much as 1000, which is far from isothermal.  It is believed that the
difference between this code and what has been done in the past with
\enzo\  stems from the Riemann solver.  The HLL family of Riemann solvers
assumes a particular wave structure in computing the interface flux.
This wave structure, for HLLC and HLLD, contains a contact
discontinuity which is not present in the isothermal Riemann fan, and
does not reduce appropriately in the $\gamma \rightarrow 1$ limit.  To combat
this, we installed the Isothermal variant of HLLD by
\citet{Mignone07}.  The results of this code on the Kim test are
nearly identical to that in figure \ref{fig.kim_ad_hlld} and not
reproduced here.  The problem seen are, of course, eliminated as
the sound speed is set as an input parameter.

\subsubsection{One-dimension MHD Caustics}\label{sec.Caustics}

This test is taken from \citet{Li06}, which initially derived from a
pure hydro version from \citet{Ryu93}. This problem is used to test
the ability of the code to 
capture shocks and to deal with hypersonic flows. 
Initially, $v_x = -\frac{\pi}{2} sin( 2 \pi x)$, $\rho=1$ and
$p=10^{-10}$. 
Caustics are formed because of the compression by the velocity field. The Mach
number of the initial peak velocity is $1.2\times 10^4$. The pressure
can easily become negative for such high Mach number flow. 
	
\begin{figure}[p]
\includegraphics[width=1.0\textwidth]{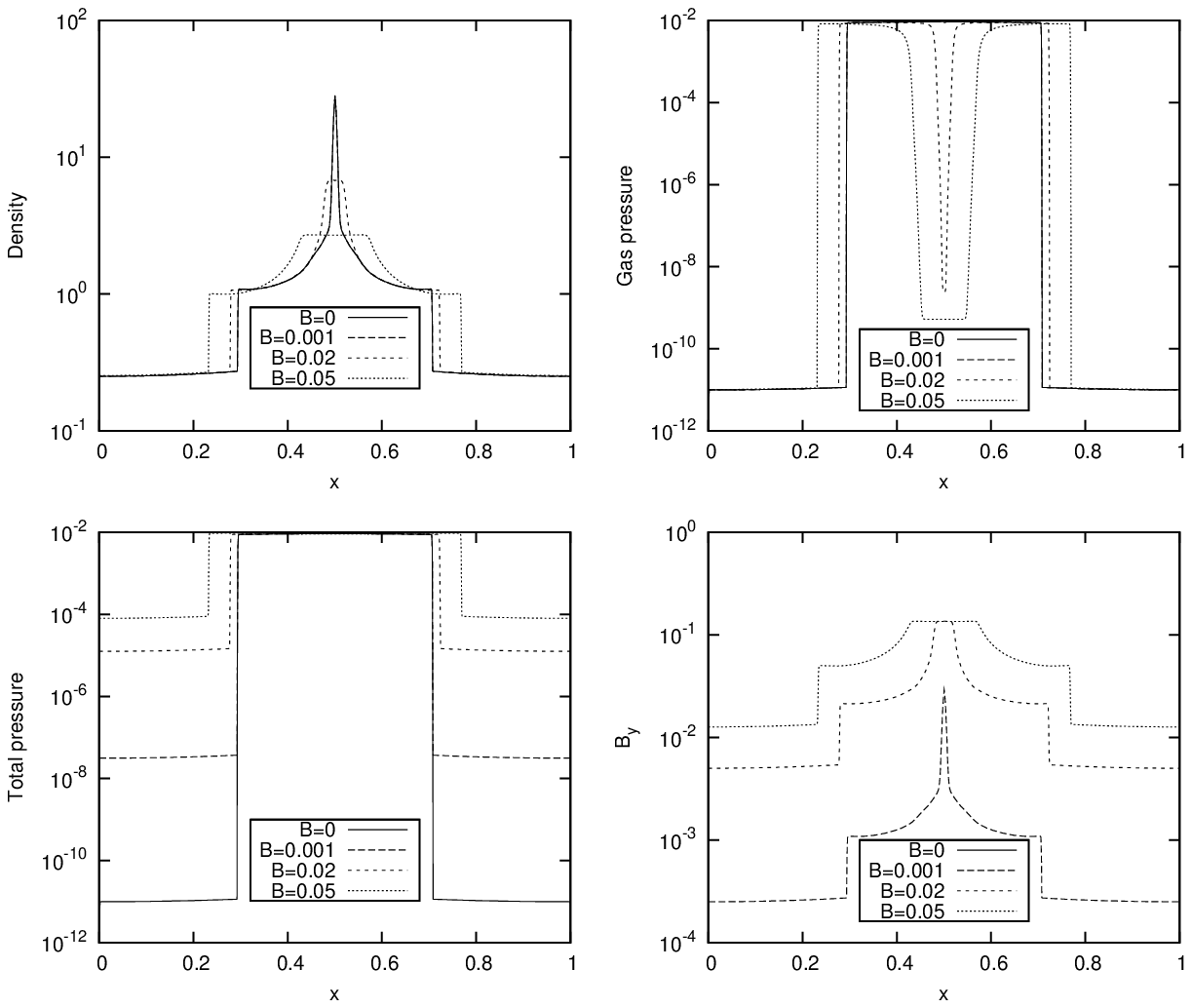}
\caption{ 1-D MHD caustics at $t=3$.  Density, gas
  pressure, total pressure and $B_y$ are
  plotted.  For the small field runs, almost no change can be seen,
  while larger field runs decrease the peak of the density
  considerably due to the increased pressure.}

\label{fig:caustic1} 
\end{figure}

We performed the test with same magnetic field settings as in
\citet{Li06}.  The magnetic field in the x and z directions are 
always zero while $B_y = 0, 0.001, 0.02$ and $0.05$. The
calculation was done with 1024 cells and the results at $t=3$
are shown in figure \ref{fig:caustic1}. Our results match the results 
from CosmoMHD \citep{Li06} quite well, as expected.

\subsubsection{The Zel'Dovich Pancake}\label{sec.ZelDovich}

\begin{figure}[p]
\includegraphics[width=1.0\textwidth]{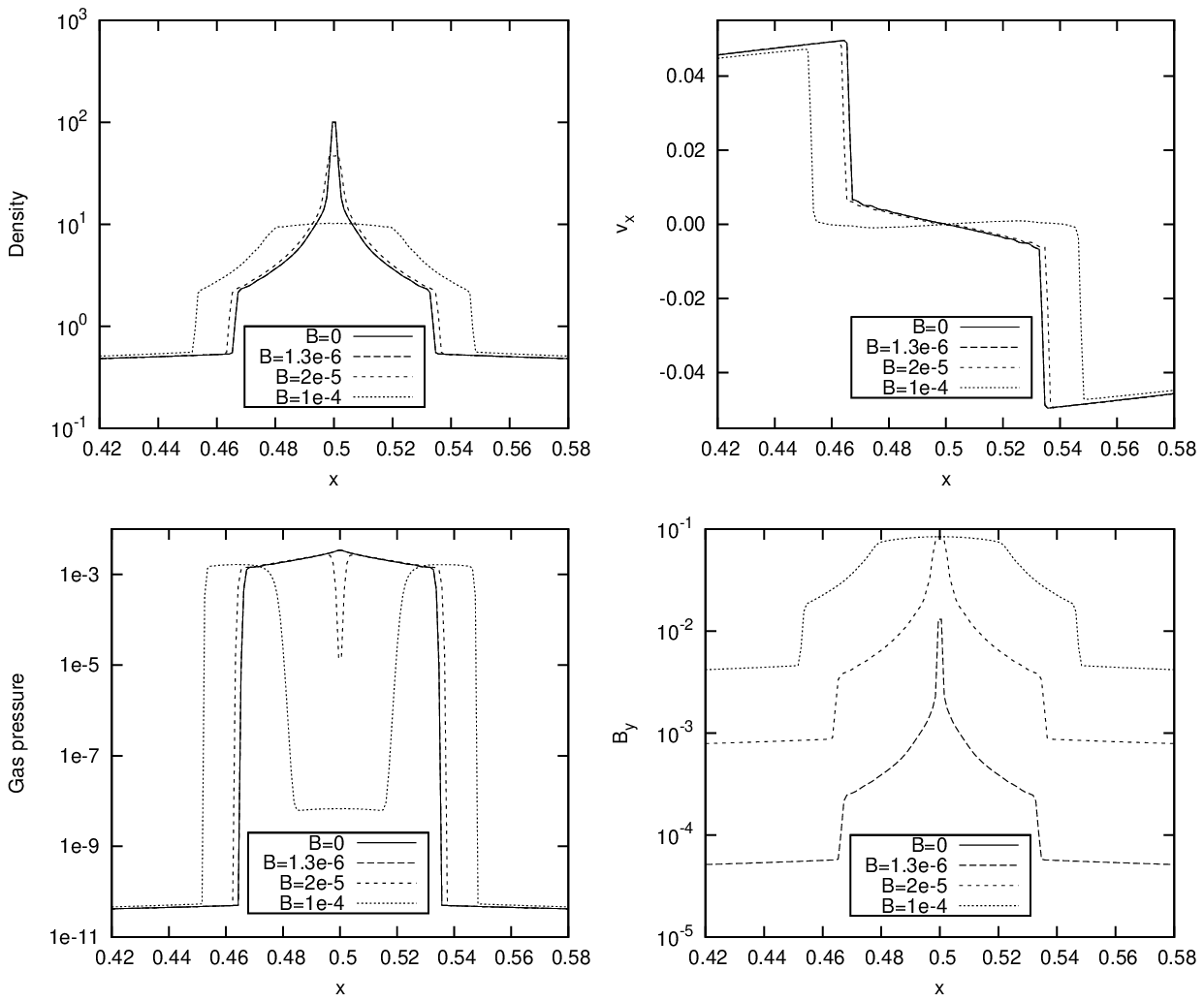}
\caption{The Zel'Dovich Pancake problem with various values of the
  magnetic field, at $t=0$.  Increasing the magnetic field strength
  increases the central magnetic pressure, reducing the density and
  changing the overal solution structure.  Results match those of \citet{Li06}.}
\label{fig:mhdzel1} 
\end{figure}

The Zel'Dovich pancake is a popular test problem for codes that
include gravity in comoving coordinates. The 
 problem setups are taken from \citet{Li06}. This takes place  in a
 purely baryonic universe with 
 $\Omega=1$ and $h=\frac{1}{2}$. The initial scale factor $a_{i}=1$
 corresponds to $z_{i}=20$. The initial velocity field is sinusoidal
 with the peak value $0.65/(1+z_{i})$, and
 $v=0$ at the center of the box. The initial comoving box size is
 $64h^{-1}Mpc$.  The shocks forms at $z=1$. The initial baryonic density and pressure are
 uniform with $\rho=1$ and $p=6.2\times 10^{-8}$.
 The tests were run with 1024 cells, both with
 and without magnetic fields. Our
results are almost identical to the results from CosmoMHD
 \citep{Li06}, as expected.  Results can be seen in figure \ref{fig:mhdzel1}.

\subsubsection{Orszag-Tang}\label{sec.OT}
The Orszag-Tang Vortex was originally developed by \citet{Orszag79} to
demonstrate that small 
scale structure can be generated by the nonlinearities in the MHD
equations.  It initially starts with a single large scale rotating velocity structure and two circular
magnetic structures.  From these simple large scale initial
conditions, substantial small scale structure is formed.  It now
serves as a standard test problem to demonstrate the 
accuracy and diffusivity of MHD codes.  

\begin{figure}[p]
\plotone{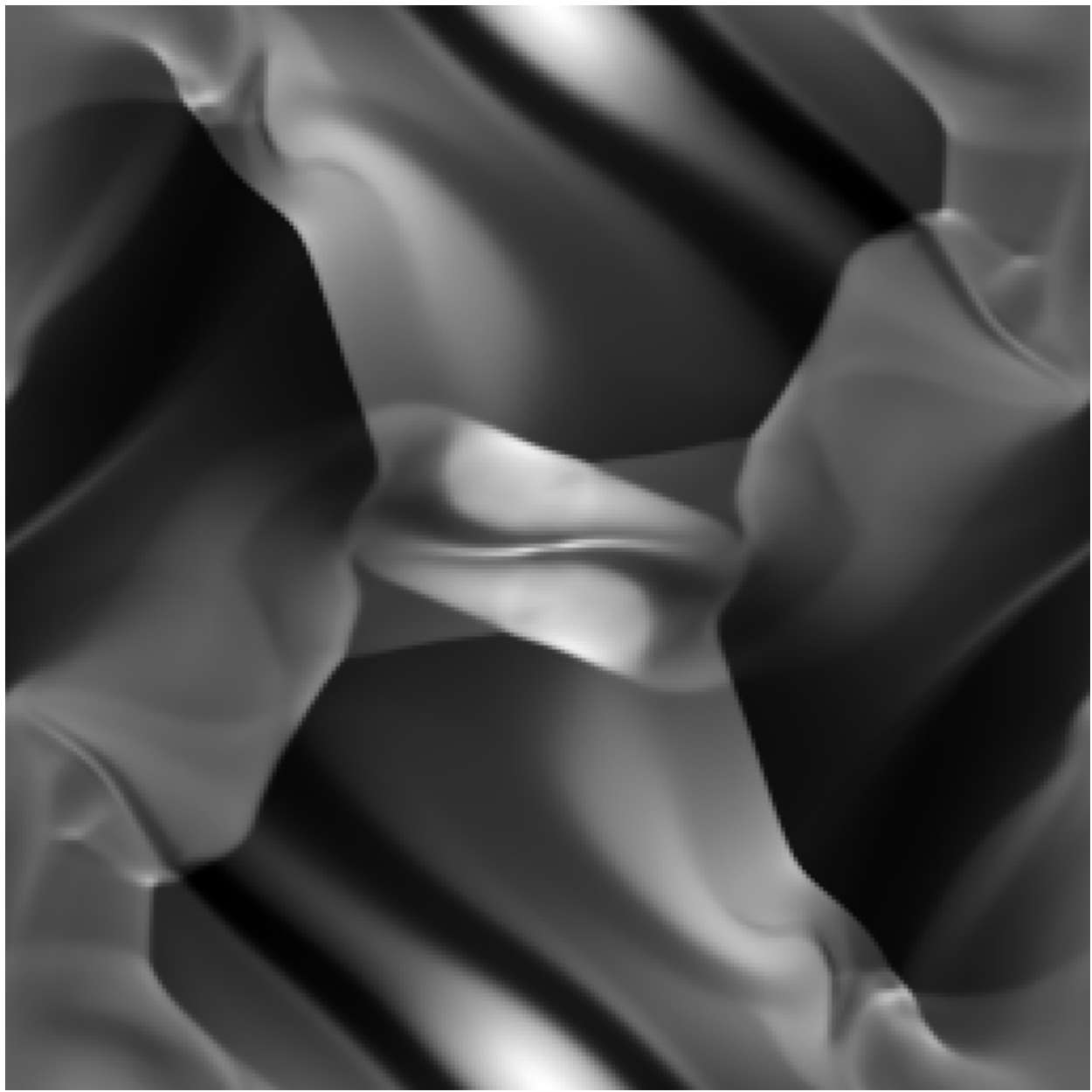}
\caption{Density from the Orszag-Tang vortex, at t=0.48.  Initial
  conditions are uniform density, with a single rotating velocity
  structure and two circular magnetic structures.  This generates
  significant small scale structure, which has been used to compare
  effective resolution of different MHD schemes.}
\label{fig.ot}
\end{figure}

The initial conditions are on a 2 dimensional periodic box, 256 zones
on a side.  $\bld{v} = v_0( -sin(2 \pi y)\hat{x} + sin( 2 \pi x)
\hat{y}, \bld{B} = B_0( -sin(2 \pi y)\hat{x} + sin(4 \pi x) \hat{y}
), v_0 = 1, B_0 = 1/\sqrt{4 \pi}, \rho_0 = 25/(36 \pi), p_0 =
5/(12\pi),$ and $ \gamma = 5/3$ which gives a peak Mach number of 1 and
peak $\beta = p_0/(B_0^2/2) = 10/3$.  Figure \ref{fig.ot}
shows the density at $t=0.48$, from which one can see that the
solution agrees with other solutions to the problem in the literature.

\subsubsection{3D Adiabatic Universe with MHD}\label{sec.Expanding}
	
We have also performed the 3D adiabatic CDM Universe test described by
\citet{Li06} both with and without magnetic fields. We also compared
the non-magnetized results with the results run using the 
PPM solver \citep{Colella84}.  Adiabatic evolution of a purely
baryonic Universe was
computed with an initial CDM power spectrum with the following
parameters: $\Omega=\Omega_{b}=1$, $h=0.5$, $n=1$ and $\sigma_{8}=1$
in a computational volume with side length $L=64h^{-1}Mpc$.  The transfer
function from \citet{Bardeen86} was used to calculate the power
spectrum of the initial density fluctuations. Evolution was done from
$z=30$ to $z=0$. We used $256^3$ cells for each simulation. The
comparisons are made at the final epoch, $z=0$.  Though this 
test is identical to that of \citet{Li06}, our results can't compared
with theirs directly since different random seeds were used for the
realization of the initial density and velocity.
	
\begin{figure}[p]
\includegraphics[width=1.0\textwidth]{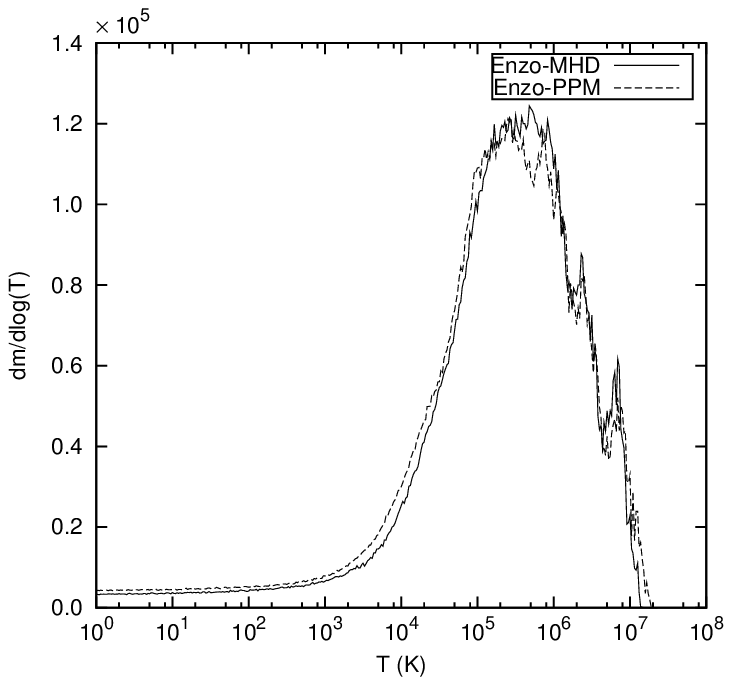}
\caption{Comparison of mass-weighted temperature histogram at $z=0$
  for the 3D purely baryonic adiabatic Universe simulation. The solid line
  is from the MHD code and the dashed line is from Enzo-PPM.} 
\label{fig:cosmo1} 
\end{figure}

\begin{figure}[p]
\includegraphics[width=1.0\textwidth]{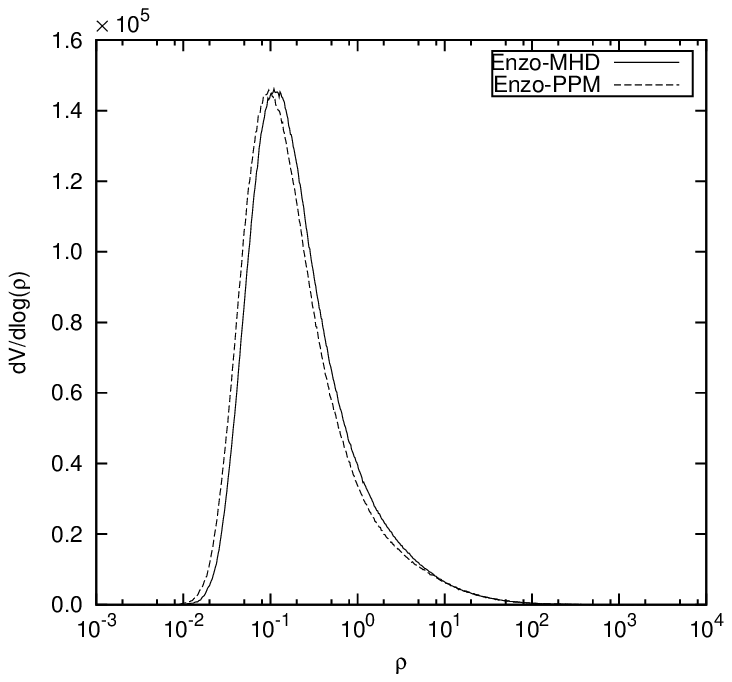}
\caption{Comparison of volume-weighted density histogram at $z=0$ for
  the 3D purely baryonic adiabatic Universe simulation. The solid line is
  from the MHD code and the dashed line is from Enzo-PPM. } 
\label{fig:cosmo5} 
\end{figure}

Figure \ref{fig:cosmo1} shows a comparison of the mass-weighted
temperature distribution, figure
\ref{fig:cosmo5} is a comparison of the 
volume-weighted density distribution.
The discrepancies between PPM and MHD
solvers are small, indicating the two codes perform roughly the same.
The nature of the differences is expected, since PPM solver has third order accuracy while
the MHD solver has second order accuracy and larger numerical diffusion.  This
allows PPM to capture shocks in fewer zones, which causes the dense shocked gas
to not only have a smaller volume fraction, but also be hotter and
slightly less dense than in the MHD solver.


\begin{figure}[p]
\includegraphics[width=1.0\textwidth]{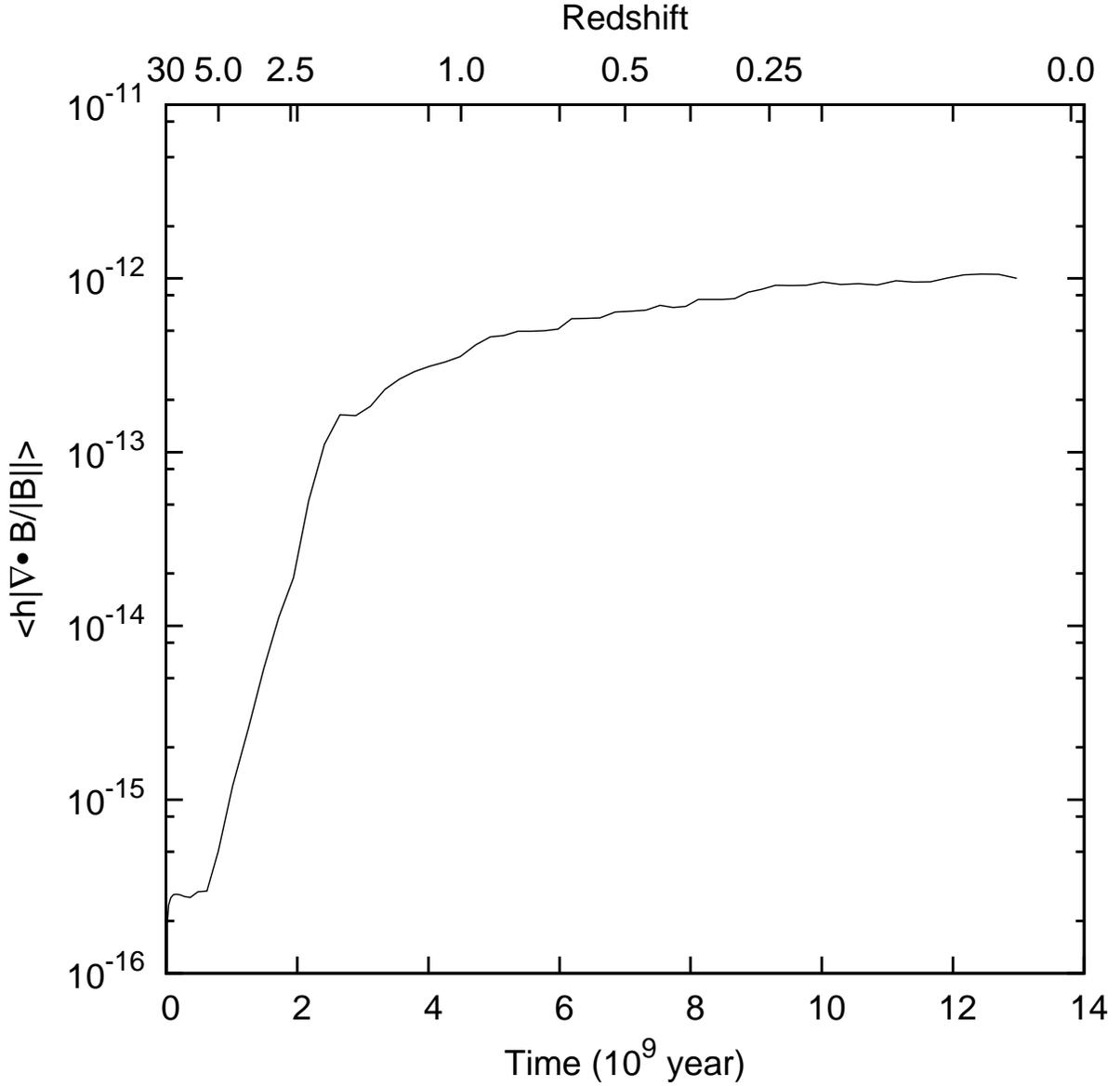}
\caption{ The scaled divergence $<|h \nabla \cdot \bf{B}/|B||>$ of the
  magnetic fields for the 3-D simulations of a purely baryonic adiabatic
  Universe.  Here $h=1/256$ is the scale length and $||B||$ is the
  local magnetic field norm, and the average is over the entire
  volume.  
Scaled divergence is a more relevant measure of
  numerical effects of divergence than the strict divergence.  As
  desired, the divergence is near the machine round off noise, the
  theoretical limit. }
\label{fig:cosmo4} 
\end{figure}

We have also done a similar run with the same initial conditions to
the above, but with an initial magnetic field, $B_x=B_z=0,~B_y=2.5
\times 10^{-9}$ Gauss, which is $4.32 \times 10^{-7}$ in code
units. 
Figure \ref{fig:cosmo4} shows the scaled divergence of the
magnetic fields, averaged over the entire box, as a function of
redshift. The scaled divergence is $<|h \nabla \cdot \bf{B}/|B||>$, where
$h = 1/256$ is the spatial scale, and $|B|$ is the local maximum
magnetic field strength, is the most relevant measure of the potential
numerical effects of divergence. The divergence of the magnetic fields is close to the
round-off error. 



\subsection {MHD Tests with AMR}
To test the Adaptive Mesh Refinement, we ran a sample of the 
tests from the previous section with AMR, to ensure no spurious artifacts are introduced
by the AMR.  These are the Adiabatic Expansion test in section \ref{sec.AdiabaticAMR}
and the one dimensional caustic and pancake tests (sections \ref{sec.CausticAMR} and \ref{sec.PancakeAMR}).

\subsubsection{Three-dimension MHD Adiabatic Expansion}\label{sec.AdiabaticAMR}

This test is taken from \citet{Bryan95}. This test uses a completely
homogenous universe with initial $T_{i}=200K$ and $v_{i}=100km/s$ in
the x-direction at an initial redshift of 
$z_{i}=20$. In the code units, the initial density is 1.0 and initial
velocity is $2.78\times 10^{-3}$ and the initial pressure is $1.24 \times
10^{-9}$. Additionally we have 
 a uniform magnetic field $B_{x}=B_{y}=B_{z}=1\times 10^{-4}$ in code 
units, which is $2.66 \times 10^{-7}G$ in cgs.

The simulation used a $16^3$ root grids with 2 levels of
refinement in the center region and ran to $z = 0$.

The expansion terms in eqns (\ref{eq.Rho}) - (\ref{eq.Induction}) operate like drag terms, so
that in the absence of a source, the velocity decreases as $v=v_{i}a^{-1}$,
the temperature as $T=T_{i}a^{-2}$ and the magnetic field should decrease as
$a^{-1/2}$.

\begin{figure}[p]
\includegraphics[width=1.0\textwidth]{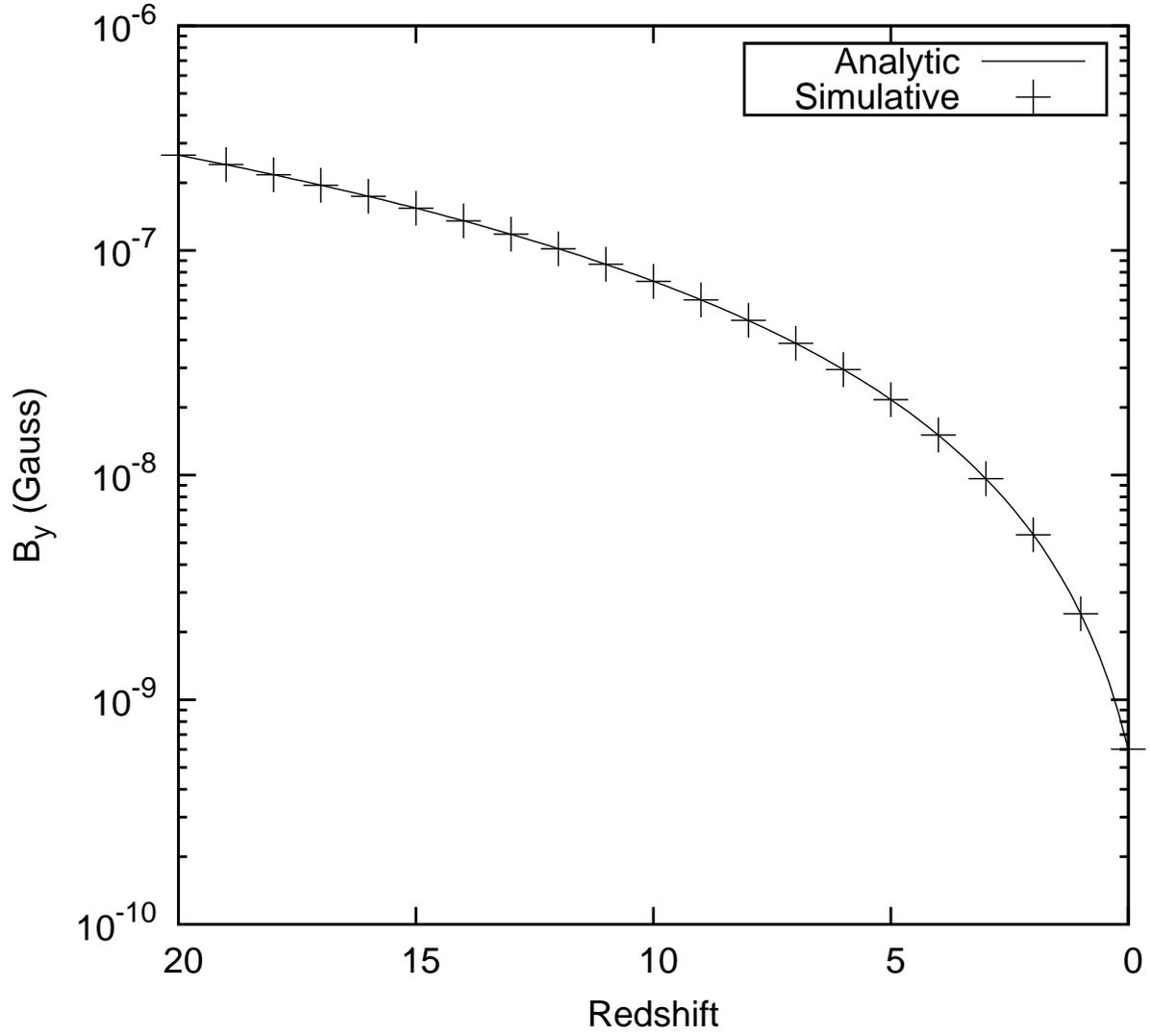}
\caption{Magnetic field in the y direction in the AMR MHD adiabatic
  expansion test.  The pluses show the results of simulation and the
  solid line is the analytic result.} 
\label{fig:adiabatic} 
\end{figure}

The temperature at $z=0$ is $0.453406K$, $0.024\%$ below the analytic
result of $0.453515K$. The velocity at $z=0$ is $4.76176km/s$, compared to
the analytic result $4.7619km/s$, a $0.0029\%$ discrepancy. The final
magnetic field strength is
$6.03 \times 10^{-10}G$ ($2.18 \times 10^{-5}$ in the code units), a difference of $0.0006\%$ 
with respect to the analytic solution.
Figure~\ref{fig:adiabatic} shows the $B_y$  as a function of redshift,
the solid line shows the theoretical value.   

\subsubsection{One-dimensional MHD Caustics with AMR}\label{sec.CausticAMR}
	
\begin{figure}[p]
\includegraphics[height=0.9\textheight]{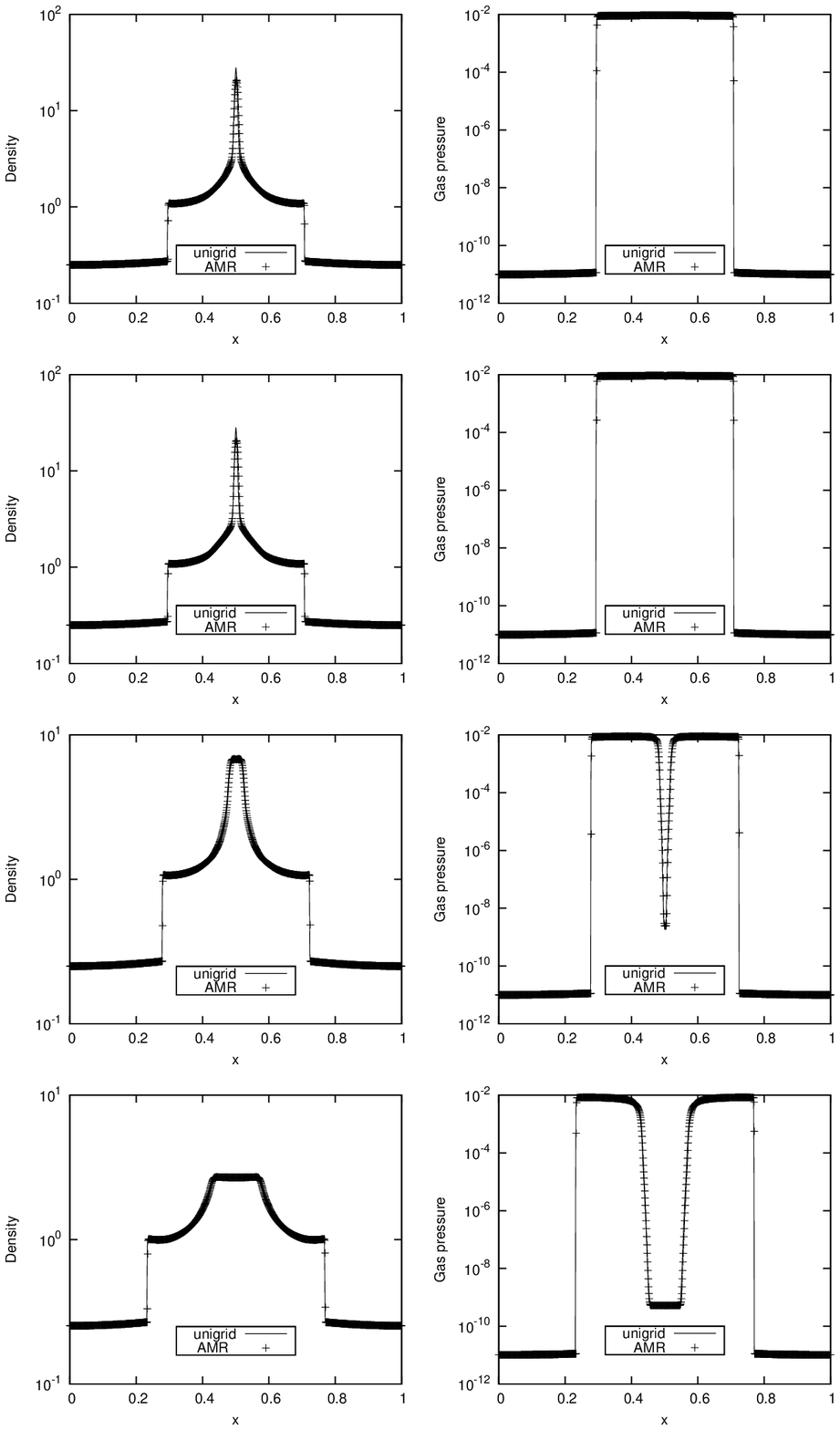}
\caption{Comparisons of density and pressure in the MHD Caustic tests,
  non-AMR vs AMR. The left column shows density and the right column shows
  gas pressure. Initial magnetic field of each row from top to bottom
  is 0, 0.001, 0.02 and 0.05.}
\label{fig:caustic2} 
\end{figure}

\begin{figure}[p]
\includegraphics[height=0.9\textheight]{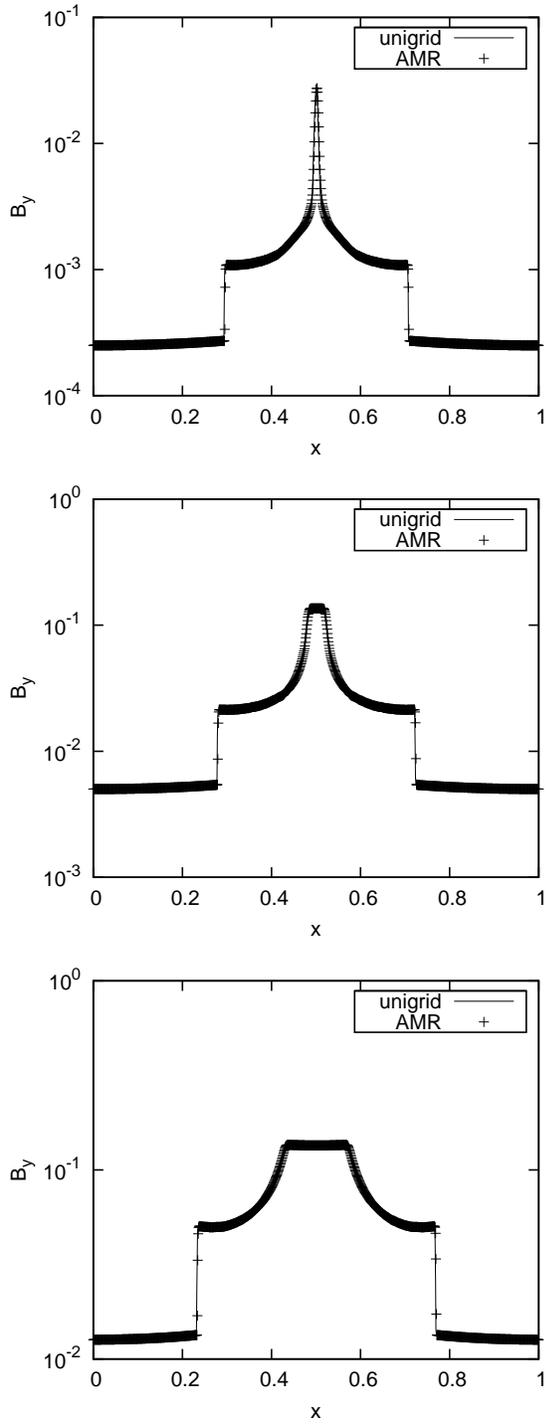}
\caption{Comparison of $B_y$ in the MHD Caustic tests, non-AMR vs AMR.
 Initial magnetic field of each panel from top to bottom is
  0.001, 0.02 and 0.05.}
\label{fig:caustic3} 
\end{figure}

We also ran the the 1d MHD Caustic test with AMR, using 256 root grid zones
with 2 levels of refinement, again by a factor of 2, giving an
effective resolution is 1024 cells. Figure \ref{fig:caustic2} shows
comparisons of density and gas pressure of 
non-AMR and AMR runs with different initial magnetic field strengths,
as described before.  Figure \ref{fig:caustic3} shows the comparisons
of $B_y$ for runs with different initial values of $B_y$. In both
plots, the AMR result is sampled to the finest resolution.  The AMR runs
give almost identical results to the unigrid runs, while the CPU time
and memory were greatly saved in the AMR runs.      
	
\subsubsection{Zel'Dovich Pancake with AMR}\label{sec.PancakeAMR}
	
\begin{figure}[p]
\includegraphics[height=0.75\textheight]{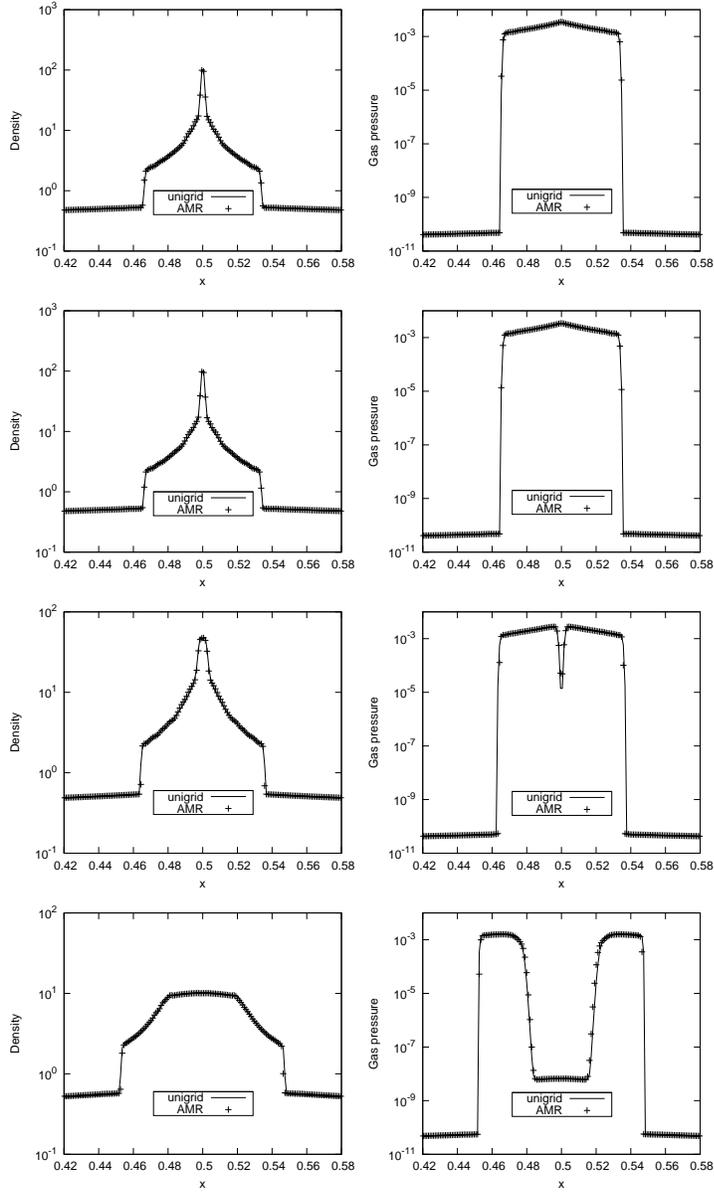}
\caption{Comparisons of density and pressure in non-AMR and AMR
  runs of the Pancake test. The left column shows density and the right column shows gas
  pressure. Initial magnetic field of each row from top to bottom is
  0, 1.3e-6G, 2e-5G and 1e-4G.}
\label{fig:mhdzel2} 
\end{figure}

\begin{figure}[p]
\includegraphics[height=0.9\textheight]{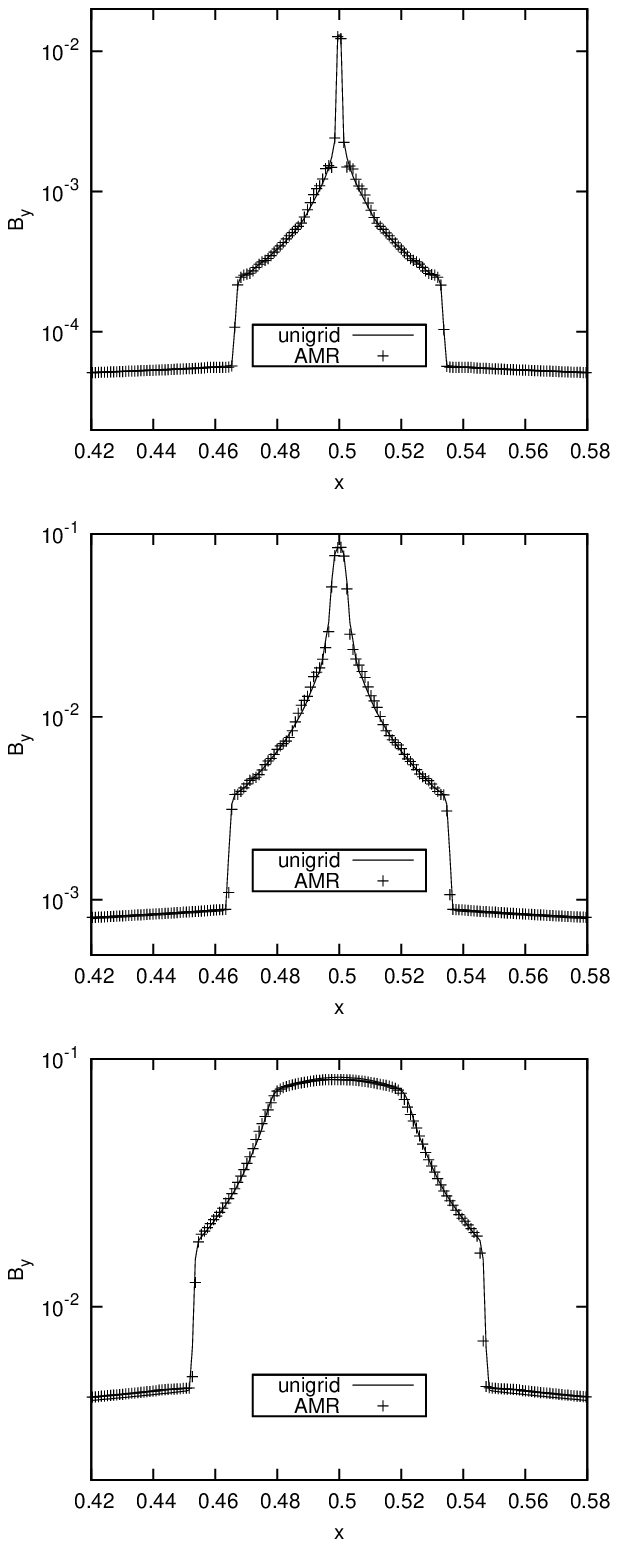}
\caption{Comparisons of magnetic y component in non-AMR and AMR
  runs of the Pancake test. Initial magnetic field of each panel from top to bottom is
  1.3e-6, 2e-5 and 1e-4G.}
\label{fig:mhdzel3} 
\end{figure}

We also ran the pancake problem with AMR.  The problem was set
up with the same initial conditions as the unigrid run, but with a
root grid of 256 root cells and 2 levels of refinement by 2.
We compared these results having effectively 1024 cells to the results
of our previous high resolution which actually had 1024 cells.
Figure \ref{fig:mhdzel2} shows comparisons of density
and gas pressure between the non-AMR and AMR runs, with different initial
values for $B_y$. Figure \ref{fig:mhdzel3} shows the comparisons
of $B_y$ with different initial values. Again, the AMR computation got
very similar results, while saving CPU and memory resources.

\subsubsection{MHD Galaxy Cluster Formations}\label{sec.Cluster}

Cluster formation (without MHD) has been studied intensively by
researchers using Enzo \citep{Norman2005,Bryan98,Loken02,Motl04,Hallman06}.
  It is one of the most important 
applications of \enzo's high dynamic range. Many cluster simulations
have been run with Enzo with a wide variety of physics (i.e. radiative
cooling, star formation, etc) and we can compare these results to
similar simulations run with MHD. More information about Enzo
simulated cluster can be found in Simulated Cluster Archive at
http://lca.ucsd.edu/data/sca/.
Here, we present just one simulation to demonstrate the MHD code.

This simulation uses a Lambda CDM
cosmology model with parameters $h=0.7$, $\Omega_{m}=0.3$, $\Omega_{b}=0.026$,
$\Omega_{\Lambda}=0.7$, $\sigma_{8}=0.928$. The survey volume is 256 $h^{-1}$ Mpc on
a side. The simulations were computed from a $128^3$ root grid with 2
level nested static grids in the center where the cluster form.  This
gives an effective resolution of $512^3$ cells
(0.5 $h^{-1}$ Mpc per cell) and dark matter particle mass 
resolution of $1.49 \times
10^{10}$ solar masses initially in the central region. Adaptive mesh
refinement is allowed only in the 
region where the galaxy cluster forms, with a total of 8 levels of
refinement beyond the root grid, for a maximum spatial resolution of
7.8125 $h^{-1} kpc$.  While the baryons are resolved at higher and
higher spatial and mass resolution at higher levels, the dark matter
particles maintain constant mass so as not to add any additional
noise. The simulations are evolved from $z=30$ to $z=0$, and all results
are shown at the redshift $z=0$. We concentrate our study on a cluster
of $M=1.2\times 10^{15}M_\odot$. 

In order to isolate the effects of the numerical approximation from
the effects of MHD, we first run the simulations adiabatically without
additional physics and the magnetic field set to zero, and compare to
a PPM run with identical parameters. In table \ref{tab:cluster1}, we list the basic parameters for the
clusters formed in each solver. The viral radius, $R_{vir}$ is
calculated for an over density $\frac{\delta\rho}{\rho}$ of
200. $M_{vir}$, $M_{dm}$ and $M_{gas}$ are the total mass, mass of the
dark matter and mass of the baryon inside the virial radius,
respectively. $T_{vir}$ is the average of the temperature of the ICM
inside the virial radius. Evidently, there is very little difference
between the results from the two solvers.

\begin{figure}[p]
\includegraphics[width=1.0\textwidth]{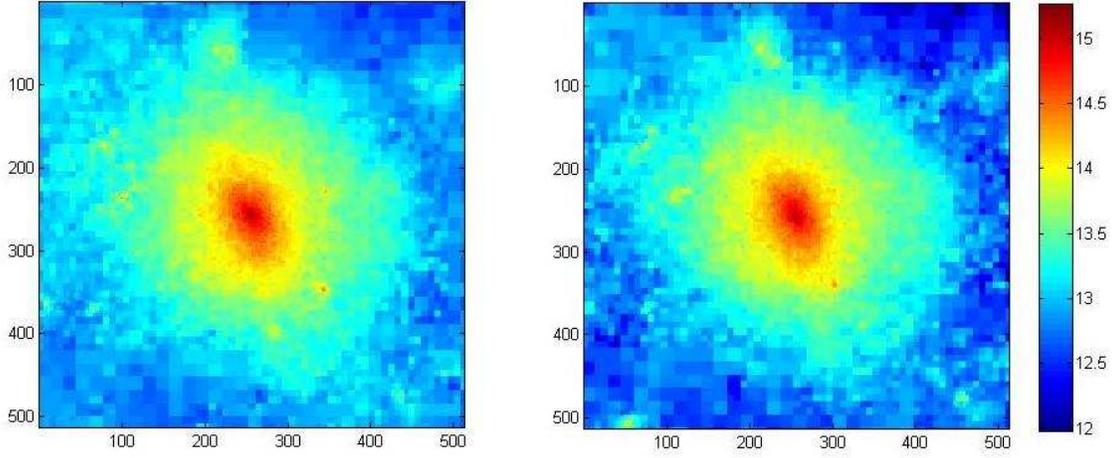}
\caption{Logarithmic projected dark matter density at $z=0$. The
  images cover the inner 4 Mpc/h of cluster centers. The left panel
  shows the result from the PPM solver and the right panel shows the result from
  the MHD solver. The color bar is in $M_\odot~Mpc^{-3}$.}  
\label{fig:cluster1}
\end{figure}

\begin{figure}[p]
\includegraphics[width=1.0\textwidth]{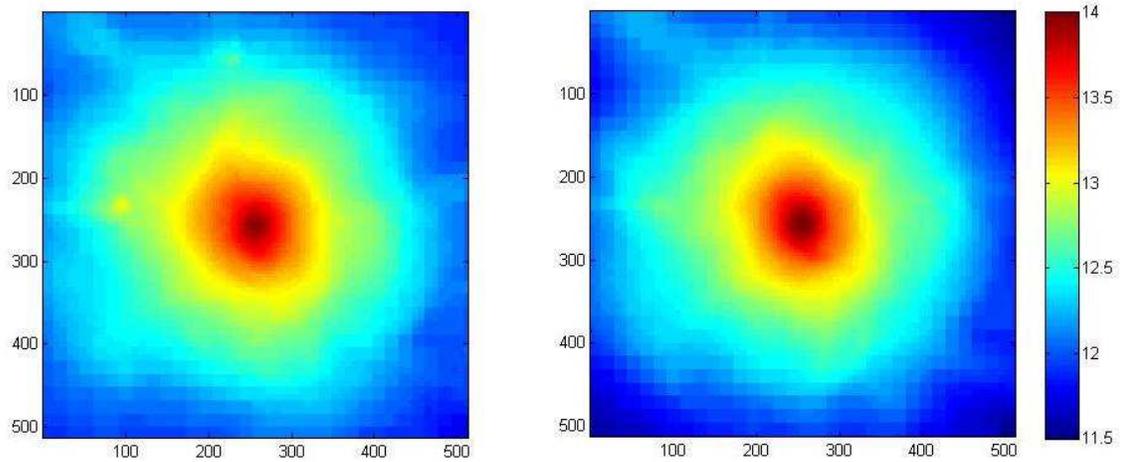}
\caption{Logarithmic projected gas density at $z=0$. The images cover
  the inner 4 Mpc/h of cluster centers. The left panel shows result
  from PPM solver and the right panel shows result from MHD
  solver. The color bar 
  is in $M_\odot~Mpc^{-3}$.}  
\label{fig:cluster2}
\end{figure}

\begin{figure}[p]
\includegraphics[width=1.0\textwidth]{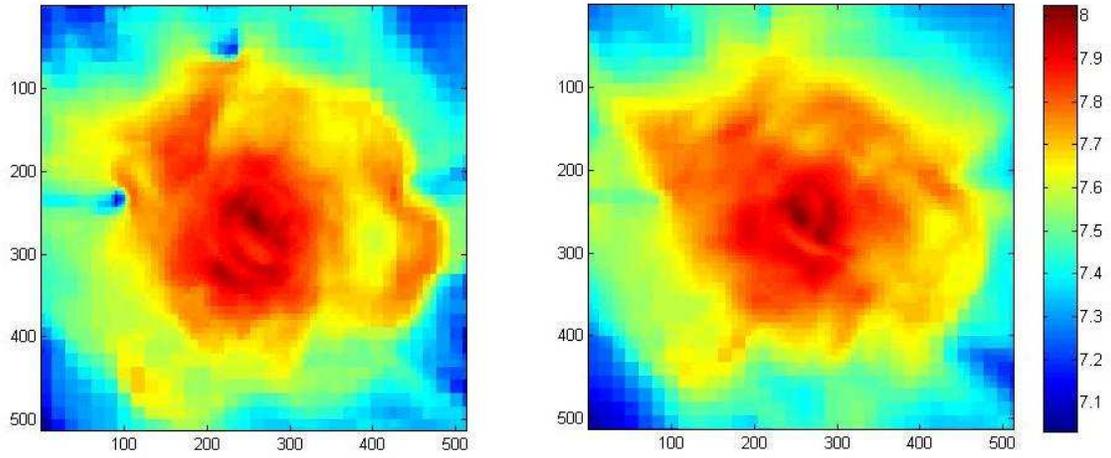}
\caption{Logarithmic projected X-ray weighted temperature at
  $z=0$. The images cover the inner 4 Mpc/h of cluster centers. The
  left panel shows result from PPM solver and the right panel shows
  result from MHD solver. The unit is Kelvin.}  
\label{fig:cluster3}
\end{figure}

Figures \ref{fig:cluster1}-\ref{fig:cluster3} show the images of the
logarithmic projections of the dark matter density, gas density, and
X-ray weighted temperature, respectively, at $z=0$. Both PPM and MHD
solvers show very similar images in all three quantities,
differing only slightly in the small scale details.

\begin{table}
\begin{center}
\caption{Cluster Properties}
\label{tab:cluster1}
\begin{tabular}{ccc}
\tableline\tableline
Parameter & Hydro PPM & MHD     \\
\tableline
$R_{vir}(Mpc)$    & 2.22946     & 2.22674  \\
$M_{vir}(M_\odot)$  &1.26462e+15  & 1.25999e+15  \\
$M_{dm}(M_\odot)$   &  1.09746e+15 & 1.09683e+15    \\
$M_{gas}(M_\odot)$  &1.67158e+14  &  1.6316e+14   \\
$T_{vir}(K)$      &8.68422e+07 &   8.66301e+07  \\
\tableline
\end{tabular}
\end{center}
\end{table}

\begin{figure}[p]
\includegraphics[width=1.0\textwidth]{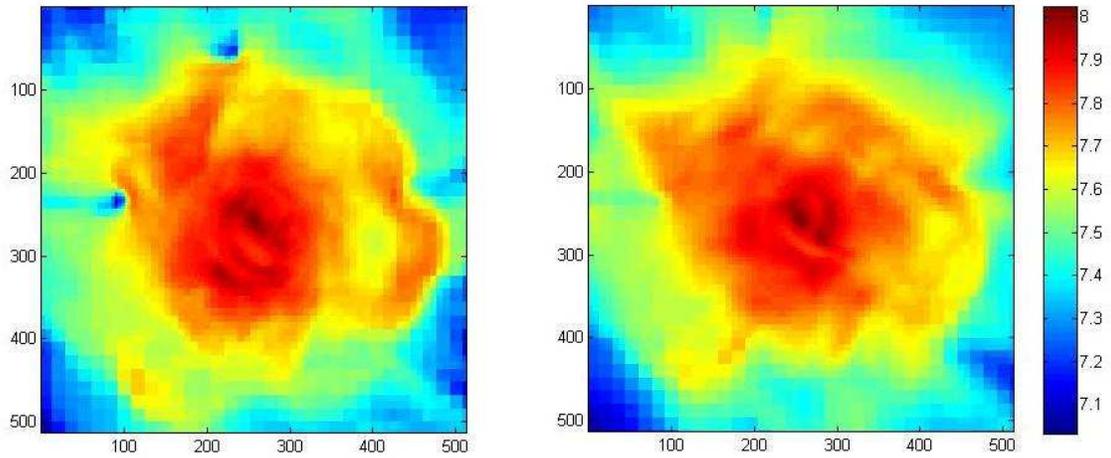}
\caption{Spherically averaged dark matter density radial profile at $z=0$ from MHD solver and PPM solver.} 
\label{fig:cluster4}
\end{figure}

\begin{figure}[p]
\includegraphics[width=1.0\textwidth]{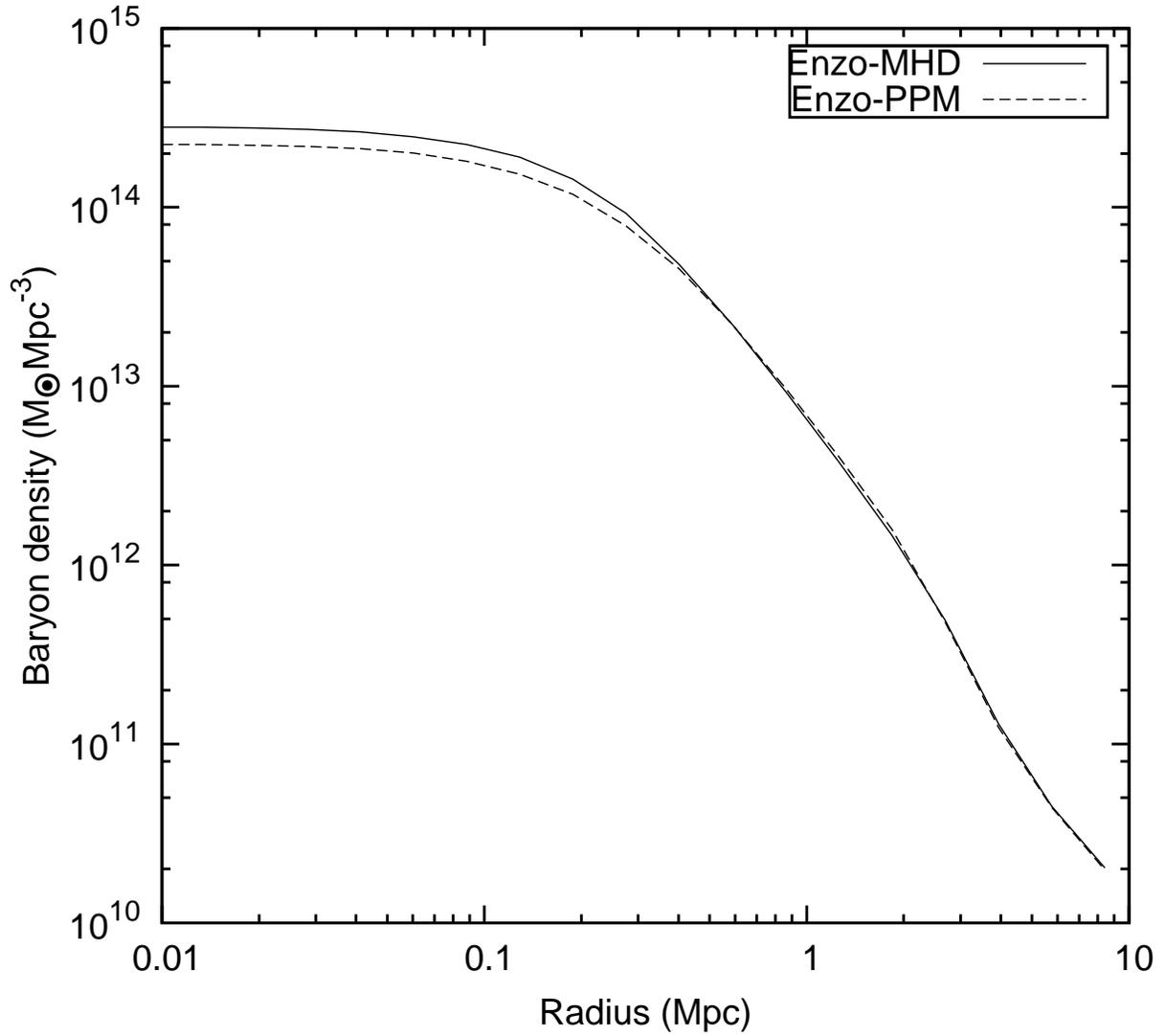}
\caption{Spherically averaged gas density radial profiles at $z=0$ from MHD solver and PPM solver.} 
\label{fig:cluster5}
\end{figure}

\begin{figure}[p]
\includegraphics[width=1.0\textwidth]{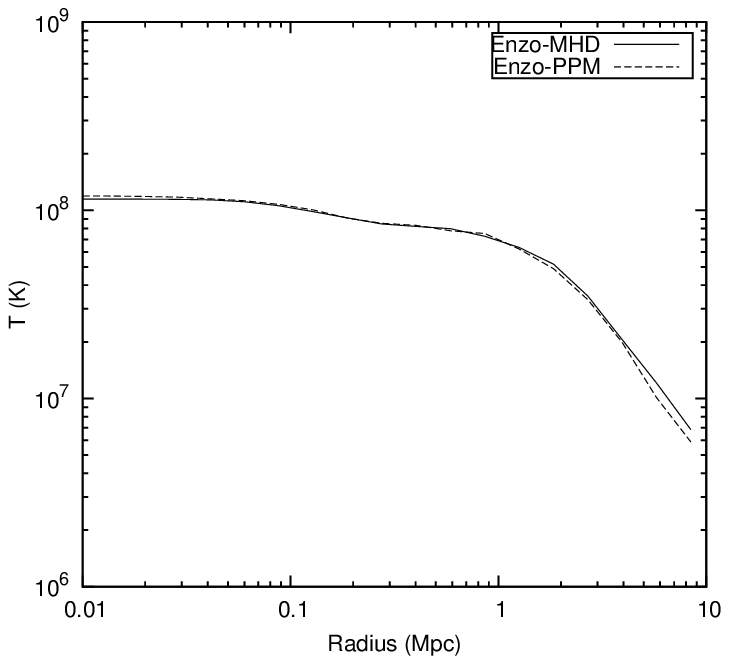}
\caption{Spherically averaged temperature radial profiles at $z=0$ from MHD solver and PPM solver..} 
\label{fig:cluster6}
\end{figure}

Figure \ref{fig:cluster4}-\ref{fig:cluster6} show the radial profiles of dark matter 
density, gas density, and x-ray weighted temperature. The profiles
match quite well in all three quantities, with only minor differences.
There is a slight deviation in the radial profiles of dark matter
density near 
the center of the cluster, but this is near the resolution limit of
the simulation, so not a trustworthy data point.  In the density
profile, it can be seen that the MHD solver gives a slightly higher
average density.   The temperature agreement is good enough to not worry about.

\begin{figure}[p]
\includegraphics[width=1.0\textwidth]{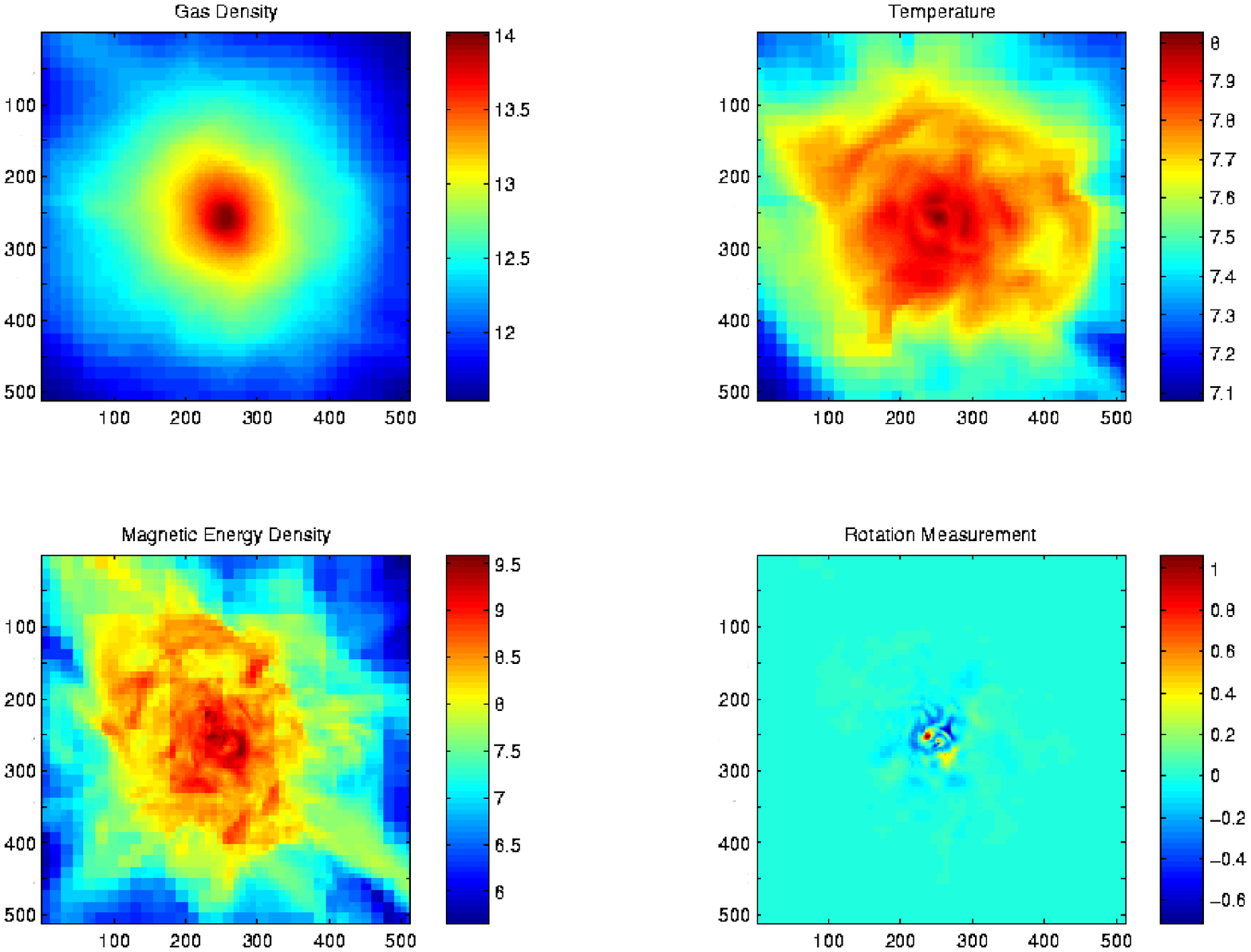}
\caption{Images of gas density ($M_\odot$ $Mpc^{-3}$), temperature (K), magnetic energy density ($erg$ $cm^{-2}$)
  and rotation measure ($rad~m^{-2}$) of the galaxy cluster simulation with an initial magnetic
  field $B_y= 9.72753 \times 10^{-10}G$.  Projections are of the inner 4Mpc/h of cluster
  center at $z=0$. }  
\label{fig:cluster7}
\end{figure}

\begin{figure}[p]
\includegraphics[height=\textheight]{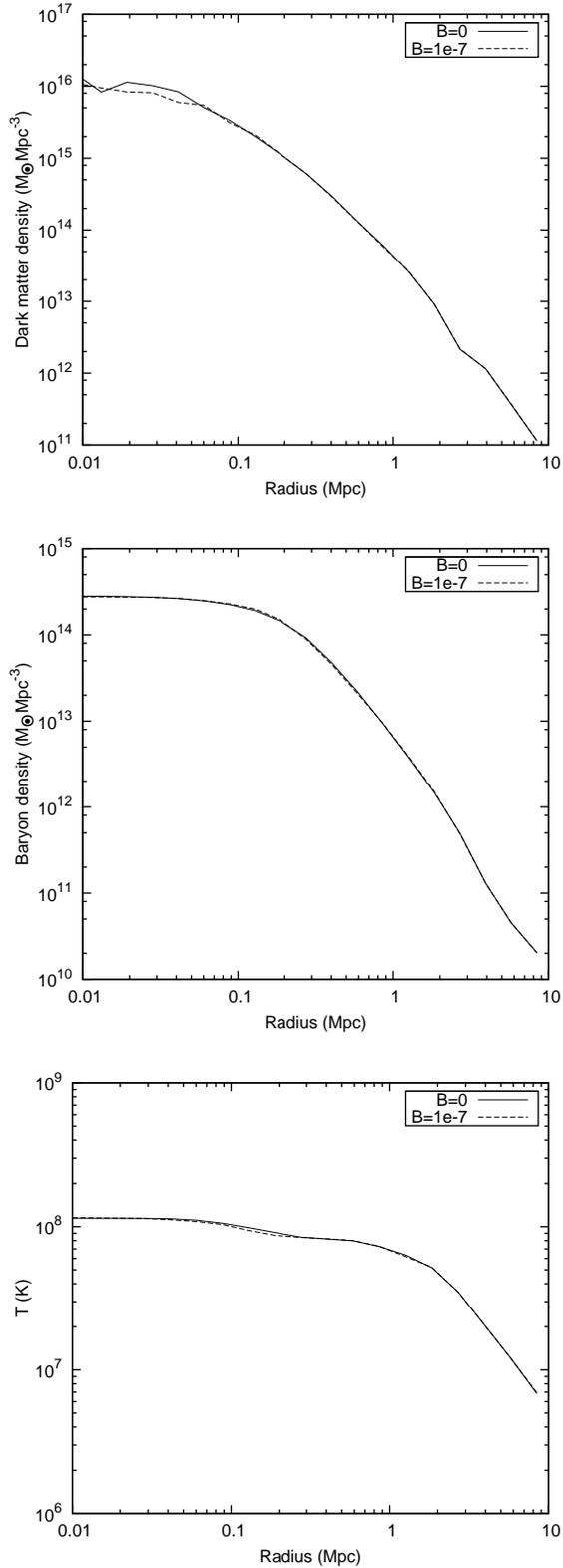}
\caption{Specially averaged radial profiles of dark matter density, baryon density and temperature of MHD simulations with zero and $B_y=
  9.72753 \times 10^{-10}G$ initial magnetic fields.}  
\label{fig:cluster8}
\end{figure}

\begin{figure}[p]
\includegraphics[width=1.0\textwidth]{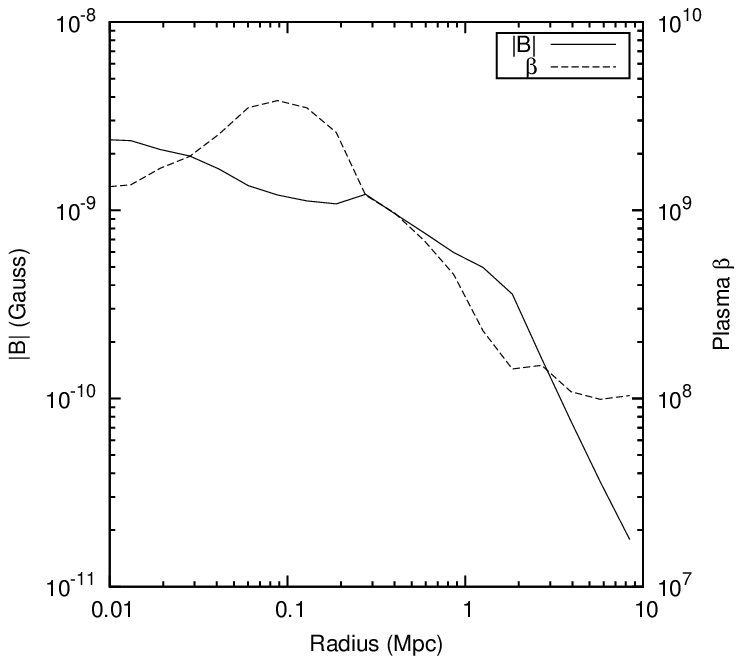}
\caption{Spherically averaged radial profiles of magnetic field strength and plasma $\beta$ of MHD simulation with $B_y=9.72753 \times 10^{10}$G initial magnetic fields.}
\label{fig:cluster10}
\end{figure}

We have also run the simulations with non-zero initial magnetic
field. A uniform initial magnetic field of $9.72753 \times 10^{-10}G$
($1\times 10^{-7}$ in code units) in the y direction was added to the
system at the start of simulation at $z=30$. Since \citet{Dolag99} has
shown that the initial magnetic fields structures are not important
to the final magnetic fields structures in their MHD SPH simulations,
no other initial magnetic fields configuration will be used in this
paper.  Figure \ref{fig:cluster7} shows 4 projections of the cluster
center: gas density, temperature, magnetic energy, and synthetic Faraday rotation
measurement $RM=\frac{e^3}{2\pi m^2 c^4}\int_{0}^{d}n_e B ds$. We can see that
the gas density and temperature images are almost identical to the MHD
run with zero magnetic fields. As expected, the magnetic energy is
concentrated in the cluster core. The maximum magnetic fields is
$1.0630270 \times 10^{-8}G$. The RM is about 2-3 $rad m^{-2}$ at the
cluster core. Figure \ref{fig:cluster8} shows comparison of the radial
profiles of the simulations with and without initial magnetic
fields, while figure \ref{fig:cluster10} depicts the volume weighted averaged radial profiles of
the magnetic field strength and plasma $\beta$.  Since $\beta$ is
quite large, these small magnetic fields acts as a passive tracer of
the plasma and has little effects on dark matter and gas dynamics.    

\begin{figure}[p]
\includegraphics[width=1.0\textwidth]{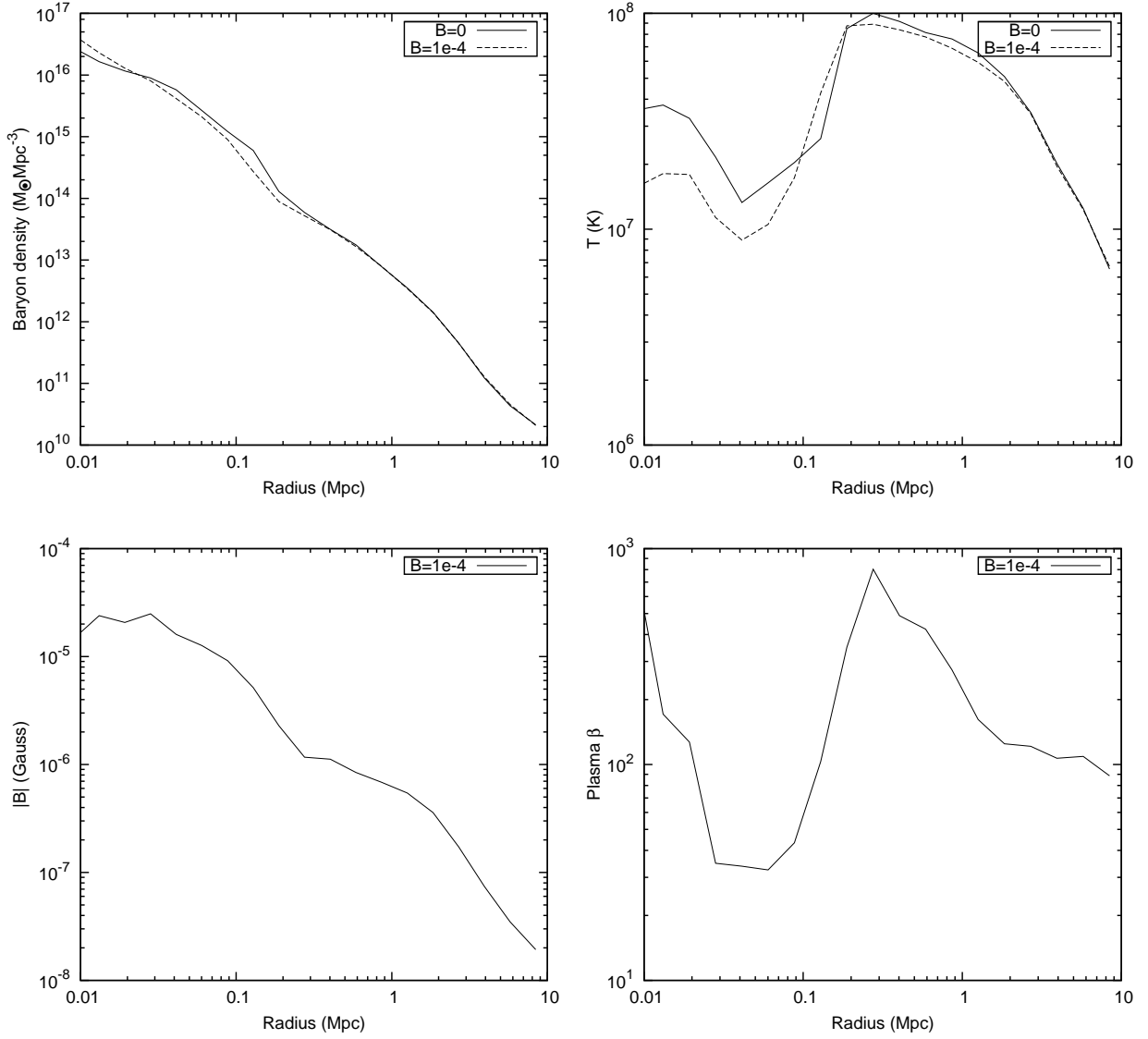}
\caption{Radial profiles of MHD simulations with zero and $B_y= 9.72753 \times 10^{-7}G$ initial magnetic fields with radiative cooling, star formation and stellar feedback.} 
\label{fig:cluster9}
\end{figure}

To further test our code, we also ran a simulation with a relatively large
initial magnetic fields. We also included radiative cooling, star
formation, and stellar feedback.  The radiative cooling
models X-ray line and bremsstrahlung emission in a
0.3 solar metallicity plasma. The star formation model
 turns cold gas into collisionless star particles at a
rate $\dot{\rho_{SF}}=\eta_{SF}
\frac{\rho_b}{max(\tau_{cool},\tau_{dyn})}$, where $\eta_{SF}$ is the
star formation efficiency factor $~0.1$, and $\tau_{cool}$ and
$\tau_{dyn}$ are the local cooling time and free fall time,
respectively. Stellar feedback returns a fraction of stars' rest energy as thermal
energy at a rate $\Gamma_{SF} = \eta_{SN}\dot{\rho}_{SF}c^2$ to the
gas. We did two runs, one without
initial magnetic fields and the other is with a large initial magnetic
fields of $B_y=1.0\times 10^{-4}$ in code units ($9.72753
\times 10^{-7} Gauss$.)  Figure \ref{fig:cluster9} shows the radial
profiles of gas density and temperature of both
runs and the magnetic field strength and the plasma $\beta$ of the run with magnetic fields.

The magnetic fields reached 20 $\mu$ G in the core region,
a few times larger than the observations
\citep{Carilli02}. In the center where $\beta$ reaches a minimum, the kinetic
energy is a few percent of the thermal
energy, as expected from \citet{Iapichino08}. The magnetic field has
become dynamically important in the cluster center.  The effect is not
significant in the density, as seen in the upper right plot in figure
\ref{fig:cluster9}, but definitely noticable in the temperature field,
as some of the thermal pressure that was balancing the collapse is
replaced by magnetic pressure.   In this way, magnetic fields may help
to cool cluster cores, giving a better match to observations.
Detailed analysis of the magnetic field structure and
their influence on the cluster will be presented in forthcoming paper.  

\section{Conclusion}\label{sec.Conclusion}

In this work, we have presented the implementation of MHD in the AMR
cosmology code \enzo\ in order to serve as a single complete reference document
for future simulations done with \menzo, and a reference for future users of the code.  
\menzo\ is capable of
multi-resolution cosmological and non-cosmological astrophysical
simulations using ideal MHD.   \enzo\ uses block structured
AMR, which solves they hydrodynamic (and now magnetohydrodynamic) PDEs
on fixed resolution patches, and communicates the finest resolution
information between coarse and fine patches in way that is
conservative in the volume-averaged quantities.  This entails 4 basic
components: the PDE patch solver, creation of fine grids (interpolation),
communication of fine data back to coarse data (projection) and
correction of the interface between coarse and fine grids (flux
correction).  MHD has the additional constraint that the divergence of
the magnetic field, \divb, must be zero at all times, which requires
additional machinery to advance the PDEs (Constrained Transport) and
some modifications to the projection and flux correction steps.  In
addition to multi-resolution hydrodynamics, \menzo\ includes the
effects of gravitational acceleration and cosmological expansion, and a
modification to the base PDE solver to account for flows with large
disparity between kinetic and thermal energies (dual energy formalism).
In \menzo, we used we use the PDE solver of
\citet{Li06} to solve the ideal MHD equations (section
\ref{sec.HyperbolicTerms}) for the patch solver, which is second
order accurate in both time and space.  We use a slightly modified version of the AMR algorithm
procedure of \citet{Balsara01} to create interpolate fine grids and project
the more accurate fine grid data to the cheaper coarse grid data (section \ref{sec.AMR} and 
appendix \ref{sec.recon}).
We have used the CT methods of \citet{Balsara99} and \citet{Gardiner05} to advance the
induction equation while maintaining the constraint \divbo (section \ref{sec.CT}.  
We have operator split the gravitational (\ref{sec.GravityTerms}) and cosmological
expansion (\ref{sec.ExpansionTerms}) terms; and included the dual
energy techniques of \citet{Ryu93} and \citet{Bryan95}. 

In section \ref{sec.Tests}, we present the results of a broad array of
tests to demonstrate the accuracy of the chosen methods.  These
include the shock tube of Brio and Wu \ref{sec.BW}, the
isothermal shock of Kim \ref{sec.Kim}, on dimensional MHD Caustics
\ref{sec.Caustics}, the famous Zel'Dovich Pancake \ref{sec.ZelDovich},
the Vortex problem of Orzag-Tang \ref{sec.OT}, an adiabatic expanding
universe \ref{sec.Expanding}.  Some of these were additionally run
with AMR, and the results compared to the unigrid case.  The results
of these overall agree with both what's been present in the literature
before and comparisons with our existing PPM solver.  As an example of the capability and
application area of this code, we present some preliminary results from a calculation
of galaxy cluster formation with magnetic fields in section \ref{sec.Cluster}

Currently underway are simulations involving protostellar
core formation, MHD Turbulence, and galaxy cluster formation and
evolution with magnetic fields.  Work has begun to include cosmic
ray acceleration, sink particles for star formation, and ambipolar
diffusion into the code.  

This work has been supported in part by NSF grants AST-0708960 AST-0808184, AST-0807768 and by NASA grant NNX08AH26G.  Additional support was supported by IGPP at Los Alamos National Laboratory.  
Simulations described in this paper were performed at the San Diego Super Computing
Center with computing time provided by NRAC allocation MCA98N0202 and LANL Institutional
HPC clusters

\section{Appendix}

\appendix				   
\section{AMR MHD Reconstruction}\label{sec.recon}

\subsection{MHD Reconstruction}\label{sec.mhdrecon}
For completeness, we will briefly outline the AMR
reconstruction used in \menzo.  The reader is encouraged to see the
details in the original paper by \citet{Balsara01}.  

In this appendix, we have dropped the subscript $f$ from the face
centered fields, as the face centered field is the only one in
question.

Balsara's reconstruction method for the magnetic field is a 3
dimensional, quadratic reconstruction of all 3 vector fields
simultaneously.  If we let ${\bf b}$ be the polynomial fit to the discrete
face centered field field $B$, the general reconstruction is

\begin{equation}
\label{eqn.recon1x}
b_x(x,y,z) = a_0 + a_x x + a_y y + a_z z + a_{xx} x^2 + a_{xy} xy +
a_{xz} xz
\end{equation}
\begin{equation}
\label{eqn.recon1y}
b_y(x,y,z) = b_0 + b_x x + b_y y + b_z z + b_{xy} xy + b_{yy}y^2 +
b_{yz} yz
\end{equation}
\begin{equation}
\label{eqn.recon1z}
b_z(x,y,z) = c_0 + c_x x + c_y y + c_z z + c_{xz} xz + c_{yz} yz +
c_{zz} z^2
\end{equation}

The coefficients are found by the following constraints:
\begin{enumerate}
  \item The analytic reconstruction should be divergence free.
  \item At the faces of the parent cell, the reconstruction should
  reduce to a bilinear reconstruction, where the slopes are
  monotonized with the minmod slope limiter.  For instance,
\begin{equation}
b_x(x=\frac{\Delta x}{2}, y) = B_{x,\iph,j,k} + \frac{\Delta_y
  B_{x,\iph}}{\Delta y} y + \frac{\Delta_z B_{x,\iph}}{\Delta z} z
\end{equation}
where
\begin{equation}
\Delta_y B_{x,\iph} = minmod( B_{x,\iph, j+1} - B_{x,\iph,j} , B_{x,\iph,j}-B_{x,\iph,j-1} ) 
\end{equation}
\begin{equation}
minmod(x,y) = 
\begin{cases}
 x, & |x| < |y|\  \text{and}\  xy > 0\\ 
y, & |y| < |x|\  \text{and}\  xy > 0 \\
0, & xy < 0
\end{cases}
\end{equation}
\end{enumerate}
The $minmod$ slope is used in order to minimize oscillations. Area
weighted averages over these polynomials are then used to assign the
fine grid values. 

Often, a fine grid patch will encroach on unrefined territory.  This
results in the refinement of coarse zones that a.)  share a face with fine
grids but b.) don't have corresponding fine grids of their
own. Balsara refers to this as ``Prolongation'' of the fine grid.   To
avoid generating any divergence at  
the boundary of the face, the interpolation polynomials need to match
the old fine data.  The interpolation equations above
(eqns \ref{eqn.recon1x} - \ref{eqn.recon1z}) do not have enough
degrees of freedom to accommodate that many data points.  In this case,
Balsara describes a new polynomial that DOES have enough degrees of freedom, by
adding $3^{rd}$ order cross terms to equations \ref{eqn.recon1x} -
\ref{eqn.recon1z}:  

\begin{align}
\label{eqn.recon2x}
b_x(x,y,z) = & a_0 + a_x x +  a_y y + a_z z + a_{xx} x^2 + a_{xy} xy +
a_{xz} xz \nonumber \\
& + a_{yz} yz + a_{xyz} xyz + a_{xxz} x^2z + a_{xxy} x^2y
\end{align}
\begin{align}
\label{eqn.recon2y}
b_y(x,y,z) = & b_0 + b_x x + b_y y + b_z z + b_{xy} xy + b_{yy}y^2 +
b_{yz} yz \nonumber \\
& + b_{xz} xz + b_{yyz} y^2 z + b_{xyz} xyz + b_{xyy} xy^2
\end{align}
\begin{align}
\label{eqn.recon2z}
b_z(x,y,z) = & c_0 + c_x x + c_y y + c_z z + c_{xz} xz + c_{yz} yz +
c_{zz} z^2 \nonumber \\
& +c_{xy}xy + c_{yzz}yz^2 + + c_{xzz}xz^2 + c_{xyz}xyz
\end{align}

The yet undetermined coefficients are found by matching the polynomial
to a bilinear fit on the face: \begin{equation}
b(x=\frac{\Delta x}{2},y,z) = B_{x,\iph,j,k} + \frac{\Delta_y
  B_{x,\iph}}{\Delta y} y +
 \frac{\Delta_z B_{x,\iph}}{\Delta z} z +
 \frac{\Delta_{yz} B_{x,\iph}}{\Delta y \Delta z} yz
p\end{equation}
and now the finite differences are taken from the finest grid:
\begin{align}
 \label{eqn.FineDyz}
 \Delta_{yz} B_{x,\iph} = 4( & ( B_{x,\iph,\jph,\kph} -
 B_{x,\iph,\jmh,\kph})  - \nonumber \\
 &( B_{x,\iph,\jph,\kmh} - B_{x,\iph,\jmh,\kmh}) ) 
\end{align}
\begin{align}
 \label{eqn.FineDy}
 \Delta_{y} B_{x,\iph} = ( & ( B_{x,\iph,\jph,\kph} -
 B_{x,\iph,\jmh,\kph})  + \nonumber \\
 &( B_{x,\iph,\jph,\kmh} - B_{x,\iph,\jmh,\kmh}) ) 
\end{align}
where $B$ is the field on the fine grid.  Note that since this is
now a centered difference, the minmod slope limiter is not used.

\subsection{Implementation in Enzo}

In order to avoid complicated book keeping routines to determine which
cells are being prolonged into, and from which direction, we formulate
only one interpolation polynomial, given by equations
\ref{eqn.recon2x}-\ref{eqn.recon2z}.  The necessary finite differences
for a given refinement region are taken from the finest data
available, as in equations \ref{eqn.FineDyz} and \ref{eqn.FineDy}.
The last four terms in each reconstruction 
polynomial are there exclusively to ensure consistency of Old Fine
Grid Data, so for faces that have no Fine Data before the reconstruction, these are set to
zero.  Since the reconstruction polynomial exactly matches the old
fine grid data, this also eliminates the need to copy the old fine
grid data to the newly refined patch.

\section{Flux Correction}\label{sec.fc}
At any given time  in an AMR simulation, there are points in space
that are described by more than one data structure.  In a finite
volume hydro calculation, with cell centered data fields, this occurs
at the boundary between coarse and 
fine grids in the Flux fields, $\Vec{F}$.  In an AMR MHD
calculation, with face centered magnetic fields, 
this occurs at the same boundary, in the face centered magnetic field,
and the edge centered electric field.  Ensuring consistency between
data is vital for the conservation of quantities like mass,
energy, momentum, and \divb.  Flux Correction is essential for this consistency.

\subsection{Conservation Form} \label{sec.consform}
It is useful to briefly describe the
basic formulation of the methods used in \enzo\ and \menzo\ before
moving on to the flux correction mechanism.

Any conservative system, such as ideal MHD, can be written in
a differential form as
\begin{equation}
 \dbd{V}{t} + \nabla \cdot F =
 0 \label{eqn.CLagain} \\
\end{equation}
where $V$ and $F$ are suitably defined, in our case by 
\ref{eqn.q} and \ref{eqn.F}.  Here we ignore any source terms.

In finite volume methods, we store average quantities of $V$ and
$F$, and re-write the conservation law in Conservation Form, using
the Fundamental Theorem and Stokes Theorem.  Starting with
eqn \ref{eqn.CLagain}, and integrating, we get:

\begin{align}
  \label{eqn.clAvg2}
  \intd{t}  \int_V \dbd{V}{t} dVdt  
   &=-\intd{t} \int_{A}  F\cdot dA dt
\end{align}
where the volume $V$ is taken from the point $(x,y,z)$ to 
$(x+\Delta x,y+\Delta y,z+\Delta z)$.
Now let
\begin{equation}
  \label{eqn.qhat}
  \hat{V}^{n} = \frac{1}{\Delta V} \int_{V} V(x,y,z,t^n) dV \\
\end{equation}
\begin{equation}
  \label{eqn.ftilde}
  \tilde{F}_{x,I+\half,J,K} = \frac{1}{\Delta y \Delta x}\int_{\Delta y, \Delta z} 
  F(x=I+\half,y,z) \cdot \hat{x} dy dz
\end{equation}  
where $\hat{x}$ is the unit vector in the $x$ direction. Similar
definitions apply  $\tilde{F}_y$ and $\tilde{F}_z$, and 
\begin{equation}
  \label{eqn.fhat}
  \hat{F}_x = \frac{1}{\Delta t} \int_{\Delta t} \tilde{F}_x dt
\end{equation}
The averaging here was taken explicitly in two steps to emphasize that
$\Delta x$,$\Delta y$ and $\Delta z$ are possibly functions of $t$, as
the are in cosmological hydrodynamics.  Putting this all together, we
get the equations in their final analytical form before discretization
(also the last form we'll be using here)
\begin{align}
  \label{eqn.consform}
  \hat{V}^{n+1}_{I,J,K} = \hat{V}^n_{I,J,K} - \Delta t (
  &\frac{1}{\Delta x}( \hat{F}_{x,I+\half,J,K} - \hat{F}_{x,I-\half,J,K} ) + \nonumber \\ 
  &\frac{1}{\Delta y}( \hat{F}_{y,I,J+\half,K} - \hat{F}_{y,I,J-\half,K} ) + \\ 
  &\frac{1}{\Delta z}( \hat{F}_{z,I,J,K+\half} - \hat{F}_{z,I,J,K-\half} ) ) \nonumber
\end{align}
Note that equation \ref{eqn.consform} is an exact equation, since only
averages and the fundamental theorem of calculus have been used up to
this point.  The trick in finite volume methods such as our MHD is
finding appropriate approximations to $\hat{F}$ that are both
accurate and stable.

\subsection{Conservation Form and AMR: Enter Flux Correction.} \label{sec.diffamr}
As mentioned at the beginning of the section, an AMR simulation has
multiple data structures representing a 
single point in space.  In entirely cell centered codes such as PPM,
the only such instance is at the surface of a fine grid boundary,
where both the fine grid and coarse grid represent the flux at that
point.  Moreover, after the fine grid field is projected into the
coarse, there's a mismatch on the coarse grid itself as to the value
of the flux at the surface.  The value of that discrepancy can be
easily found.  After the projection, a coarse grid at a point
$(I,J)$ has the value (restricting to 2d, for clarity) 
\begin{equation}
 \hat{V}^{n+1}_{I,J} = \sum_{ \substack{i =I\pm \quart \\ j = J\pm
 \quart}} \hat{q}^{n+1}_{i,j}
\end{equation}
where lower case quantities denote the value of the fine grid data.  Expanding
the time update for $\hat{q}^{n+1}$ in space and time, we find that
\begin{align}
 \hat{V}^{n+1}_{I,J} = \sum_{ 
 \substack{i =I\pm \quart \\ j = J\pm \quart}} 
 \hat{q}^{n}_{i,j} - & (\sum_{m=n}^{n+1}\sum_{x,j=J\pm \quart} \frac{\Delta t^m}{\Delta V^m}\hat{f}^m_{I+\half,j} + 
& - \sum_{m=n}^{n+1}\sum_{x,j=J\pm \quart} \frac{\Delta t^m}{\Delta
   V^m}\hat{f}^m_{I-\half,j}) \\
 - (y~and~z~terms)\nonumber
\end{align}
By construction of the interpolation polynomial (and 
projection at the last timesteps) the first term is just equal to
$\hat{V}^n_{I,J}$, which means that, by equation \ref{eqn.consform}
$\hat{V}_{I,J}$ effectively sees, at the point $I+\half$, 
\begin{equation}
  \label{eqn.fluxdiff}
  \frac{\Delta t}{\Delta V}\hat{F}_x = \sum_{m=n}^{n+1}\sum_{x,j=J\pm \quart}
 \frac{\Delta t^m}{\Delta V^m}\hat{f}^m_{I+\half,j} := <f_x>
\end{equation}
However, for the cell $(I-1,J)$, which has no corresponding fine grid
flux, $\hat{F}_{I+\half}$ come from the discretization method on the
coarse grid.  There is
absolutely no reason for the two to match, so we have a discrepancy in
the descriptions of the data.  This can be solved by simply replacing
the less refined data that $\hat{V}_{I+1,J}$ used with the more
refined average, given by equation \ref{eqn.fluxdiff}:
\begin{equation}
  \hat{V}_{I+1,J, \fclabel} = \hat{V}_{I+1,J} + \frac{\Delta t}{\Delta
  V}\hat{F}_{x,I+\half,J} - \sum_{m} \sum_{j} \frac{\Delta t^m}{\Delta
  V^m} \hat{f}^m_{x,I+\half,j}
\end{equation}
Now every place $\hat{F}_{x,I,J}$ show up in our method, the exact same
approximation is used.  

\subsection{Flux Correction and MHD}\label{sec.mhdfc}
A similar formalism to that described in \ref{sec.consform} is used for
to advance the magnetic fields in \menzo, but instead of using volume 
averages, we use area averages.  The magnetic evolution is given by
the induction equation:
\begin{equation}
  \label{eqn.induction}
  \dbd{ \Vec{B} }{t} = - \curl \Vec{E}
\end{equation}
When discretized, equation \ref{eqn.induction} yields the equation 
\begin{align}
 \label{eqn.inductiond}
 \hat{B}^{n+1}_{x,I+\half,J} = \hat{B}^n_{x,I+\half,J} - \frac{\Delta
 t}{\Delta y \Delta z} 
 (& \Delta z (\hat{E}_{z,I+\half,J+\half,K} -
 \hat{E}_{z,I+\half,J-\half,K} ) +\\ 
 & \Delta y (\hat{E}_{y,I+\half,J,K+\half} -
 \hat{E}_{y,I+\half,J,K-\half} ) )  \nonumber
\end{align}
where
\begin{equation}
  \label{eqn.bhat}
  \hat{B}^n_{x,I+\half,J,K} = \frac{1}{\Delta y \Delta z} \int_A 
  \Vec{B}(x=I+\half,y,z,t^n)\cdot \hat{x} dy dz
\end{equation}
\begin{equation}
  \label{eqn.ehat}
  \hat{E}^n = \frac{1}{\Delta t} \int_t^{t+\Delta t} \frac{1}{\Delta
  x}\int_{x} \Vec{E} \cdot dl dt
\end{equation}
which is also exact, and the main problem is finding a suitable
approximation for $\hat{E}$.

Again, after the area-weighted projection of the fine grid field $\hat{b}_x$
into the coarse grid $\hat{B}_x$, there's a discrepancy
between the electric field at a refined point on the surface of a
refined grid, as it's seen by both grids that have subgrids and grids
that don't.
In \cite{Balsara01}, he suggests a similar flux 
correction mechanism to that of the standard hydro, described in \ref{sec.diffamr}.  However, due to
an issue with the initial implementation of flux correction in \enzo\ (which
has since been fixed) and ease of computational logic, we chose a different route.  In \menzo, instead
of projecting fine grid magnetic fields into coarse magnetic fields and then correcting zones in the coarse
grid, we project the electric field and then take the curl of the entire
coarse grid.  Thus, all coarse grid magnetic fields see the most
accurate data at the same time, and no a-posteriori correction needs
to be done.  Where there are no subgrids, the coarse grid sees an
electric field that comes from the CT module in section \ref{sec.CT},
and where there are subgrids it sees
\begin{align}
  \label{eqn.eproj}
  \hat{E}^n_{z,\imh,\jmh,k} = 
   \frac{\Delta t^n}{\Delta t} & (e^{n+\half}_{z,\imh,\jmh,k-\quart} + 
    e^{n+\half}_{z,\imh,\jmh,k+\quart}) +  \nonumber \\
   \frac{\Delta t^{n+\half} }{\Delta t} &(e^{n+\tquart}_{z,\imh,\jmh,k-\quart} + 
     e^{n+\tquart}_{z,\imh,\jmh,k+\quart})
\end{align}
While a complete flux correction treatment would potentially save on
memory and flops, in practice the extra memory is negligible
compared to the total memory and time used by the rest of \enzo, and
the extra floating point operations done here are offset by increase
cache utilization of the data, as the entire grid is done in a single
stride one sweep instead of an essentially random access pattern.

As described in section \ref{sec.BoundaryConditions}, some of the
subgrids get their boundary conditions updated from the parent zones.
Because of this, the curl of the magnetic field is actually taken
twice.  The first time is done immediately after the hyperbolic
update, in order to ensure that the parent zones are up to date for
the interpolation of the ghost zones of the subgrids that need it.
The second time is after the subgrids project their electric field to
the parent, to ensure maximal accuracy of the parent grids.  This
additional call takes negligible time, as the curl has relatively few operations. See 
appendix \ref{sec.Schematic} for the details of this order of operations.

\section{Schematic for the Cosmological MHD Code}\label{sec.Schematic}

In this section, we present a schematic of the MHD code, for clarity
and easy reference.

Step 0.-- We start with conserved quantities 
density, total energy, and momentum ($\rho_{BM}^{n}, 
E_{total}^n, \mathbf{p}_{DM}^n)$, and primitive quantities velocity and
gas pressure ($v_{BM}^{n}$, $P^{n}_{gas}$) for the
baryonic matter; face and cell centered magnetic fields ($B_{c}^{n}$,
$B_{f}^n)$; and Lagrangian dark matter mass, position, and velocity
($\rho_{DM}^n, \mathbf{x}^{n}, \mathbf{v}_{DM}^{n})$.  These are all
at time $t^n$.  Where needed, primitive quantities will be described
by $U = (\rho_{DM},P_{gas},\mathbf{v}_{DM}, \mathbf{B})$, and conserved
quantities by $V = (\rho_{DM},E_{total},\mathbf{p}_{DM}, \mathbf{B}).$
Conversion between the two is done as needed.

Step 1. Solve Poisson's equation for the acceleration field at $t^{n+\half}$
\begin{align}
   \phi^{n} \Longleftarrow & \rho_{BM}^{n}+\rho_{DM}^{n}   \\ 
   \phi^{n+1/2}  = & \phi^{n}(1+\frac{\Delta t^{n}}{2\Delta t^{n-1}})- \phi^{n-1}\frac{\Delta t^{n}}{2 \Delta t^{n-1}}  \\
    g_{i}^{n+1/2} = & \frac{1}{2a^{n+1/2}\delta x_{i}}(\phi_{i+1}^{n+1/2}-\phi_{i-1}^{n+1/2}) 
\end{align}

Step 2.-- Update particle positions and velocities.  (Strictly
speaking, this happens after the Expansion step, but the narrative
works better if it's here.)
\begin{eqnarray}
  \bld{v}_{DM}^{n+1/2} & = & \bld{v}_{DM}^{n} - \frac{\Delta t^{n}}{2}\frac{\dot{a}^{n+1/2}}{a^{n+1/2}}\bld{v}_{DM}^{n} - \frac{\Delta{t}^{n}}{2}\bld{g}^{n+1/2}  \\
  \bld{x}_{DM}^{n+1} & = & \bld{x}_{DM}^{n} + \Delta t^{n}(\bld{v}_{i,DM}^{n+1/2}/a^{n+1/2})  \\
    v_{i,DM}^{n+1} & = & v_{i,DM}^{n+1/2} -  - \frac{\Delta t^{n}}{2}\frac{\dot{a}^{n+1/2}}{a^{n+1/2}}v_{i,DM}^{n+1/2} - \frac{\Delta{t}^{n}}{2}g_{i}^{n+1/2} 
\end{eqnarray}

Step 3.-- Apply half of the gravitational and expansion update to
the fields that require it, to obtain the temporary state
$\tilde{U} = (\rho, \tilde{P}_{total}^n,\tilde{\mathbf{v}}_{BM}^{n}, \tilde{B}_{c}^{n})$
\begin{eqnarray}
   \tilde{\mathbf{v}}_{BM}^n  & = & \mathbf{v}_{BM}^{n} - \frac{\Delta
     t^{n}}{2}\frac{\dot{a}^{n}}{a^{n}}
   \mathbf{v}_{BM}^{n}-\frac{\Delta t^{n}}{2}\frac{1}{a^{n}}\mathbf{g}^{n+1/2} \\
   \tilde{p}^{n} & = & p^{n} - \frac{\Delta t^{n}}{2}\frac{2\dot{a}^{n}}{a^{n}} p^{n} \\
   \tilde{\mathbf{B}}_{c}^{n} & = &  \mathbf{B}_{c}^n - \frac{\delta
     t^{n}}{4} \frac{\dot{a}^{n}}{a^{n}} \mathbf{B}_{c} \\
   \tilde{U} & = &  (\rho, \tilde{P}_{total}^n,\tilde{\mathbf{v}}_{BM}^{n}, \tilde{B}_{c}^{n})
\end{eqnarray}  

Step 4. Compute interface states at $i\pm \half, n + \half$ using
linear spatial reconstruction and second order time integration:
\begin{eqnarray}
U_{i+\half,L}^{n+\half}, U_{i+\half,R}^{n+\half} \Longleftarrow \tilde{U}_{i-1}, \tilde{U}_{i}, \tilde{U}_{i+1}, \tilde{U}_{i+2} 
\end{eqnarray}

Step 5. Compute approximation of the flux in equation \ref{eqn.F} at
the interface $i+\half$.   This is done by solving the Riemann problem
using one of the solvers mentioned in section  \ref{sec.HyperbolicTerms}
\begin{align}
\hat{F}^{n_\half}_{i+\half} = Riemann(U^{n+\half}_{i+\half,L},U^{n+\half}_{i+\half, R})
\end{align}

Step 6. Update the conserved quantities with the new fluxes:
\begin{align}
  (V^{n+1}_i)_\mhdlabel = V^{n}_i -  \frac{\Delta t}{\Delta x}
  [\hat{F}_{i+\half}-\hat{F}_{i-\half} ]
\end{align}

Step 7.  Compute Electric field from Fluxes
\begin{align}
  E_{i+\half,j+\half}^{n+\half} \Longleftarrow \hat{F}_{i+\half}
\end{align}

Step 9.  Update magnetic fields from electric fields for the first time.  
\begin{eqnarray}
  B_f^{n+1} = B_f^{n} - \frac{\Delta t}{a} \nabla \times E^{n+\half}_{i+\half,j+\half}
\end{eqnarray}

Step 8.--Gravitational step for the baryonic matter, with time centered density
\begin{equation}
   (\mathbf{p}_{i,BM}^{n+1})_{\mhdlabel, \gravlabel} = (\mathbf{p}_{i,BM}^{n+1})_\mhdlabel - \Delta
   t^{n}\frac{(\rho^n + \rho^{n+1}_\mhdlabel)}{2}g_{i}^{n+1/2}
\end{equation}

Step 9.--Expansion step for the baryonic matter,
\begin{eqnarray}
    (\mathbf{v}_{BM}^{n+1})_{\mhdlabel, \gravlabel, \explabel} & = & \frac{1 - (\Delta
      t^{n}/2)(\dot{a}^{n+1/2}/a^{n+1/2})}{1+(\Delta
      t^{n}/2)(\dot{a}^{n+1/2}/a^{n+1/2})}(\mathbf{v}_{BM}^{n+1})_{\mhdlabel
      \gravlabel}\\
    p^{n+1} & = & \frac{1 - (\Delta t^{n})(\dot{a}^{n+1/2}/a^{n+1/2})}{1+(\Delta t^{n})(\dot{a}^{n+1/2}/a^{n+1/2})} (p^{n+1})_\mhdlabel
\end{eqnarray}

Step 10. Recurse to finer grids.  Integrate fine grids from $t^n$ to $t^{n+1}$
\begin{eqnarray}
V_{FineGrids}^{n+1} \Longleftarrow V_{FineGrids}^n
\end{eqnarray}

Step 11.--Flux correction step for conserved baryon field quantities
\begin{eqnarray}
  V^{n+1}_{\mhdlabel, \gravlabel,\explabel, \fclabel} \Longleftarrow
  (\hat{F}^{n+1/2}),(\hat{F}^{n+1/2})_{FineGrids}, V^{n+1}_{\mhdlabel \gravlabel, \explabel} 
\end{eqnarray}

Step 12.--Project conserved baryon field quantities and electric field
from fine grids to coarse grids.  This is done \emph{after} the flux correction step to avoid any
bookkeeping errors.  The average is taken over $\Delta t^n$ and the surface of each $FineGrid$.
\begin{eqnarray}
V^{n+1}_{ParentGrid} = & <V^{n+1}_{FineGrid}>_{t,surface} \\
E^{n+\half}_{ParentGrid} = & <E^{n+\half}_{FineGrid}>_{t,surface}
\end{eqnarray}

Step 13.  Update magnetic fields from electric fields for the final time.
\begin{eqnarray}
  B_f^{n+1} = B_f^{n} - \frac{\Delta t}{a} \nabla \times E^{n+\half}_{ParentGrid}
\end{eqnarray}

Step 14. Apply expansion to the Face Centered Fields
\begin{align}
    B_{f,\explabel}^{n+1} = &  \frac{1 - (\Delta t^{n}/4)(\dot{a}^{n+1/2}/a^{n+1/2})}
    {1+(\Delta t^{n}/4)(\dot{a}^{n+1/2}/a^{n+1/2})} 
    (B_{f}^{n+1})
\end{align}

Step 15. Compute cell centered magnetic field from face centered (with
the expansion subscript from step 9 dropped for clarity)
\begin{align}
  B_{c,x,i,j,k}^{n+1} = 0.5*( B_{f,x,i+\half,j,k} + B_{f,x,i-\half,j,k} ) \nonumber\\
  B_{c,y,i,j,k}^{n+1} = 0.5*( B_{f,y,i,j+\half,k} + B_{f,y,i,j-\half,k} ) \\
  B_{c,z,i,j,k}^{n+1} = 0.5*( B_{f,z,i,j,k+\half} + B_{f,z,i,j,k-\half} )\nonumber
\end{align}

Step 16. We have now finished an update of this level.  Rebuild the hierarchy from this level down.
\begin{align}
  V^{n+1}_{New~FineGrids} \Longleftarrow &  V^{n+1} \\
  B^{n+1}_{f, New~FineGrids} \Longleftarrow &  B^{n+1}_{f}
\end{align}

\bibliographystyle{apj}
\bibliography{apj-jour,ms}  

\end{document}